\documentclass[12pt]{article}
\usepackage{hyperref}
\usepackage{cite}
\usepackage{graphicx}
\usepackage{subfigure}
\usepackage[top=3cm, bottom=3cm, left=1.5cm, right=1.5cm]{geometry}
\usepackage{amsmath,amsfonts,amssymb,mathtools}
\usepackage{graphicx}
\usepackage{float}
\usepackage{breqn}
\usepackage{multicol}
\usepackage{multirow}
\usepackage{booktabs}
\usepackage{array,makecell, cellspace}
\begin{document}
	\setcounter{page}{1}
	
	\pagestyle{plain} \vspace{1cm}
	\begin{center}
\Large{\bf Dissipative Quintessential Cosmic Inflation}\\
\small \vspace{1cm}{\bf Kourosh Nozari\footnote{knozari@umz.ac.ir}},\quad {\bf Fateme Rajabi \footnote{f.rajabi@umz.ac.ir}}\quad and \quad {\bf Narges Rashidi\footnote{ n.rashidi@umz.ac.ir}}\\
\vspace{0.25cm}
Department of Theoretical Physics, Faculty of Basic Sciences,\\
University of Mazandaran,\\
P. O. Box 47416-95447, Babolsar, IRAN
	\end{center}
	
\vspace{1cm}
\begin{abstract}	
In this paper we construct a dissipative quintessential cosmic inflation.
For this purpose, we add a multiplicative dissipative term in the standard
quintessence field Lagrangian. We consider the specific form of
dissipation as the time integral including the Hubble parameter
and an arbitrary function that describes the dissipative properties of the quintessential scalar field.
Inflation parameters and observables are calculated under slow-roll approximations and a
detailed calculation of the cosmological perturbations is performed in this setup.
We consider different forms of potentials and calculate
the scalar spectral index and tensor-to-scalar ratio for a constant as well as
variable dissipation function. To check the reliability of this model, a numerical analysis
on the model parameters space is done in confrontation with recent observational data.
By comparing the results with observational joint datasets at $68\%$ and 95$\%$ confidence levels,
we obtain some constraints on the model parameters space, specially the dissipation factor
with e-folds numbers $N=55$ and $N=60$. As some specific results,
we show that the power-law potential with a constant dissipation factor and $N=60$ is mildly consistent with
observational data in some restricted domains of the model parameter space
with very small and negative dissipation factor and a negligible tensor-to-scalar ratio.
But this case with $N=55$ is consistent with observation considerably.
For power-law potential and variable dissipation factor as $Q=\alpha \phi^{n}$, the consistency with observation is
also considerable with a reliable tensor-to-scalar ratio. The quadratic and quartic potentials with variable
dissipation function as $Q=\alpha \phi^{n}$ are consistent with Planck2018 TT, TE, EE+lowE+lensing
data at the $68\%$ and $95\%$ levels of confidence for some intervals of the parameter $n$.\\


\end{abstract}
\newpage
	
\section{Introduction}

Observations of the distant type Ia Supernovae~\cite{Rie98,Per99} and Cosmic Microwave Background Radiation (CMB)~\cite{Hin13,Ade16,Agh,Ade2016}
have confirmed that our universe is currently in a phase of positively accelerated expansion.
A cosmic fluid composed of ordinary baryonic matter and obeying the perfect fluid equation
of state is not capable to explain the positively accelerated expansion of the universe~\cite{Sen01}.
Also, observations have revealed that less than $5\%$ of the matter content
of the universe is ordinary, baryonic matter, while more than $95\%$ of the total content
of the universe \emph{probably} consists of yet unseen and mysterious dark matter and dark energy~\cite{Bla09,Cai09}.
We stress on ``probability" since it is possible in essence and equally well to attribute this cosmic speed up to
the modification of the gravitational (geometrical) part of the field equations on vast scales;
the issue of modified gravity~\cite{Nojiri2007,DeFelice2010} and also braneworld gravity~\cite{Maartens2010}.

The simplest form of dark energy to explain the late time accelerated expansion is a non-zero
cosmological constant correspond to a positive vacuum energy density. But it suffers from some
theoretical and phenomenological problems such as unknown origin, lake of dynamics and
also need for a huge amount of fine-tuning~\cite{Cop6,Stei9}. Hence, several approaches have been proposed to solve the cosmological
constant problem. In this regard, one way is to introduce a dynamical dark energy component that is capable in essence to avoid
the extreme fine-tuning of the cosmological constant~\cite{Ratra,Wetterich,Caldwell,Caldwell1,Caldwell2,Padmanabhan1,Sen,Sen1}.
This dynamical component could be a canonical scalar field; a quintessence, with an equation of state parameter $-1<w_{\phi}<-\frac{1}{3}$,
a quintom field~\cite{Cai09} as a superposition of a quintessence and a phantom field, or even some
more involved fields such as tachyon fields~\cite{Nozari2} (and references therein).
On the other hand, the cosmic inflation is actually a very fast positively accelerating phase of expansion of the early
universe that lasted for a very short time. This early time cosmic inflation can solve/address the
problems of the standard big bang model such as the horizon, flatness and relics problems~\cite{Gu81, Lin90, Lid96, Bas18, Her17, Yan16, Mar14}.
In addition and even more importantly, the density fluctuations during cosmic inflation provides the seeds for the formation of the large scale structure in
the universe~\cite{Mar92}, imprint of which are seen as anisotropy in CMB spectrum. In a simple inflation model, a canonical scalar field called inflaton
rolls down its potential and drives the desired amount of inflation~\cite{Lin83,Kal09,Cre14}.
Cosmological inflation is a viable theory that describes almost correctly the early universe
in agreement with the recent observations~\cite{Ade16}.

Quintessence models have been proposed as alternative theories to cosmological
constant in order to explain the late time cosmic speed up~\cite{Pee8,Fuj3,Far4,Pee3}.
In this approach, the cosmological evolution is modeled by a
canonical scalar field $\phi$ in the presence of a self-interaction potential. A large number of literature
have been devoted to explore various aspects of this proposal in last two decades.
Just as some instance, the dynamics of a quintessence in the presence of non-relativistic matter has been studied in details for
many different potentials in Refs.~\cite{Cop98,Cal98,Zla99,Ng01,Cor03,Lin06}. The nonminimal
coupling of the quintessence field and dark matter was considered in Ref.~\cite{Ame0}. The Hubble
tension can be reduced by considering a quintessence field with transition from a matter-like
to a cosmologically constant behavior between recombination and the present epoch~\cite{Val19}.
For some other studies on quintessence theories see Refs.~\cite{Nozari2b,Tsu3,Bar9,Cho0,The1,Adi2}.

To solve the problems attributed to quintessence model, such as the coincidence problem in
the context of initial conditions~\cite{Dim22,Dim21}, one way is to link it to an
inflationary scenario called \textit{quintessential inflation} which was firstly
suggested by Peebles and Vilenkin~\cite{Pee9} (see also some related literature in Refs.
~\cite{Pel9,Sen2,Dim01,Dim02,Kag1,Gio03,Sam04,Yah2,Ros5,Bas9,Hos4,Hos14,Har1}).
In this scenario, one uses a unified theoretical framework to describe inflation and
dark energy. A quintessence can play the roles of both inflaton and
dark energy, in the early and late time cosmic history respectively. Quintessence can
generate inflation in the early universe if it rolls down on a sufficiently flat potential
~\cite{Dim7}. Since the quintessence field must survive until today to reproduce
the current cosmic acceleration, it cannot collapse at the end of inflation.
Therefore, quintessence inflaton does not oscillate at the bottom of the
potential, but rolls down to the quintessential plateau. Hence, the universe
must be reheated through a mechanism other than the decay of the
quintessence inflaton field at the end of quintessential inflation. At the end of inflation, the scalar field
enters the kinetic regime. Therefore, reheating of the universe in a
quintessential inflation occurs through a phase transition of the universe
from inflation to \emph{kination} where the adiabatic regime is broken, allowing
the creation of particles~\cite{Joy97}. The post inflationary particle production
occurs through a variety of mechanisms, including preheating~\cite{Cam03,Dim18},
curvaton reheating~\cite{Fen03,San07,Mat07}, gravitational reheating~\cite{Chu09},
Ricci reheating~\cite{Dimo18,Opf19} and warm quintessential inflation~\cite{Dim19,Ros19,Gan21}.

Dissipative systems were firstly studied in the thermodynamics and statistical physics.
Also, dissipation occurs in quantum mechanics which is a consequence of the
dissipative interactions of a quantum system with its environment~\cite{Cal3,Raz5}.
Dissipation appears also at macroscopic level such as viscous and frictional forces
which are modeled by internal variables that evolve in an irreversible
way~\cite{Mai7,Luo0}. These theories have been extensively investigated in
astrophysical and cosmological contexts, and play an important role in the early time
evolution of the Universe~\cite{Chi0,Pun8,Yan9}. Dissipation can generally be
attributed to the interaction of a given physical system with an external (e.g. thermal)
bath, or to interaction with another physical system. Dissipative scalar field theories
play a significant role in various problems in physics, for example, in Casimir physics
a scalar field is an oscillating field that interacts linearly with some external matter
field defined inside or on some specific interaction surface~\cite{Dal11}. Various forms
of the dissipative Klein-Gordon equation have been investigated mostly in the framework
of warm inflationary cosmological models~\cite{Ber5,Bel01,Ber99,Ber06,Zha09,Santos2023,Berera2023,Kamali2023}.

In this paper, we explore some aspects of a dissipative quintessential inflation.
We consider a Lagrangian description of a dissipative canonical scalar field, based on a
variational principle, which is inspired by a simple damped harmonic oscillator. We
investigate the dissipative phenomena in this quintessential inflation by
adding a multiplicative dissipative term in the standard Lagrangian of the quintessence scalar field.
This multiplicative dissipative term is expressed in an exponential form as usual.
Recently Harko in a seminal work has shown that a dissipative quintessence field model provides an effective
dynamical possibility to replace the cosmological constant and also to explain recent
cosmological observational data~\cite{Har23}. Here we firstly study the equation of motion using
Euler-Lagrange equations and then deduce the physical properties and basic
characteristics of dissipation systems. In this case, the Klein-Gordon equation
depends on the dissipation exponent and function. Also, we obtain the energy-momentum
tensor of the dissipative scalar field through the variational principle which depends on
the dissipative function. Then we study the effects of dissipation on the inflationary
dynamics of the model. One of the main motivations in conducting the present study is
to see the role that dissipation plays throughout the cosmic dynamics, from the early inflationary
epoch toward the late time dark energy driven positively accelerated expansion. We study also the initial
cosmological perturbations in this setup as an important probe to see the
viability and feasibility of a dissipative quintessential inflation.
We examine different forms of the dissipation coefficients in this setup to see
the effects of dissipation on inflationary dynamics of the universe and related observables.
Depending on the scalar field potential model and dissipation function, we
face with different slow roll parameters. By confronting the calculated values
of inflation parameters/observables with recent observational data, the viability of this cosmic inflation
model is checked.

This paper is structured as follows. In Section 2, we present the Lagrangian of
quintessence model by considering the dissipative term and drive the basic equations
that we will need in the rest of the paper. In section 3, we discuss the cosmological
perturbation theory. In section 4, we study dynamical inflation with power-law  and
exponential potential for constant as well as variable dissipative function. We derive
the slow roll parameters and examine our generalized inflationary model in
confrontation with the recently observational data and find constraints on
the parameters space of the model. Section 5 is devoted to conclusion. \\

\section{The setup}

We consider a dissipative extension of the quintessence field, in which
the standard Lagrangian of the quintessence field is multiplied by an
exponential function of the type $e^{\Gamma(g_{\alpha \beta},\phi,x^\alpha)}$. In this regard, we consider a
gravitational model with a non-minimally coupled dissipative quintessence
field, described by the Lagrangian density $L_{\phi}$, and an ordinary matter
term with Lagrangian density $L_m$, which the action is as follows
\begin{eqnarray}\label{eq1}
S=\int{\Big[\frac{1}{2\kappa^2}R+L_{\phi}+L_m\Big]\sqrt{-g}d^4x}\,.
\end{eqnarray}
We express the action of dissipative scalar field in the general covariant form as~\cite{Har23}
\begin{eqnarray}\label{eq2}
S_\phi=\int{L_{\phi}\sqrt{-g}d^4x}=\int{e^{\Gamma(g_{\alpha \beta},\phi,x^\alpha)}\bigg[-\frac{1}{2}g^{\mu\nu}{\triangledown_{\mu}\phi}{\triangledown_{\nu}\phi}-V(\phi)\bigg]\sqrt{-g}d^4x}\,,
\end{eqnarray}
where the dissipation exponent ${\Gamma(g_{\alpha \beta},\phi,x^\alpha)}$ is an
arbitrary scalar function of the metric tensor, the scalar field and the coordinates.
$V(\phi)$ is the potential of the scalar field, $\phi$. By varying action \eqref{eq1}
with respect to the metric, we obtain the gravitational field equations in the presence
of a dissipative quintessence field
\begin{eqnarray}
R_{\mu\nu}-\frac{1}{2}g_{\mu\nu}R=\kappa^2\Big(T_{\mu\nu}^{(\phi)}+T_{\mu\nu}^{(m)}\Big)\,,
\end{eqnarray}
where $T_{\mu\nu}^{(m)}$ is the ordinary matter energy-momentum tensor which is defined as
\begin{eqnarray}
T_{\mu\nu}^{(m)}=\frac{2}{\sqrt{-g}}\frac{\delta \sqrt{-g}L_m}{\delta g^{\mu\nu}}\,.
\end{eqnarray}
By variation of action \eqref{eq2} with respect to the scalar field $\phi$, we obtain
the covariant Klein-Gordon equation as follows
\begin{eqnarray}\label{eq5}
\Box\phi+g^{\mu\nu}{\triangledown_{\mu}\phi}{\triangledown_
\nu{\Gamma(g_{\alpha \beta},\phi,x^\alpha)}}-\frac{d V(\phi)}{d \phi}-\frac{\partial{\Gamma(g_{\alpha \beta},\phi,x^\alpha)}}
{\partial\phi}\Big(-\frac{1}{2}g^{\mu\nu}{\triangledown_\mu \phi}
{\triangledown_\nu \phi}+V(\phi)\Big)=0\,.
\end{eqnarray}
The energy-momentum tensor of the scalar field is
\begin{equation}
T_{\mu \nu}^{(\phi)}=2\frac{\delta L_{\phi}}{\delta g^{\mu \nu}}-L_{\phi}g_{\mu \nu}\,,
\end{equation}
and for the dissipative scalar field, we obtain
\begin{eqnarray}
T_{\mu \nu}^{(\phi)}=e^{\Gamma(g_{\mu \nu},\phi,x^\mu)}\bigg[{\triangledown_{\mu}\phi}
{\triangledown_{\nu}\phi}+\big(\Theta_{\mu \nu}-g_{\mu \nu}\big)\Big(\frac{1}{2}g^{\alpha \beta}
{\triangledown_{\alpha}\phi}{\triangledown_{\beta}\phi}+V(\phi)\Big)\bigg]\,,
\end{eqnarray}
where $\Theta_{\mu \nu}$ is defined as
\begin{equation}
\Theta_{\mu \nu}(g_{\alpha \beta},\phi,x^{\alpha})=2\frac{\delta {\Gamma
(g_{\alpha \beta},\phi,x^\alpha)}}{\delta g^{\mu \nu}}\,.
\end{equation}
Note that for $\Theta_{\mu\nu}=0$ or $\Gamma=0$, we recover the energy-momentum
tensor of the standard non-dissipative quintessence field. Inspired by the standard
form of the energy-momentum tensor of a perfect fluid, the energy-momentum
tensor of the dissipative scalar field can be written as follows \cite{Har23}
\begin{eqnarray}
T_{\mu\nu}^{(\phi)}=e^{\Gamma(g_{\mu \nu},\phi,x^\mu)}\Big[(\rho_\phi+p_\phi)u_{\mu}
u_{\nu}+(\Theta_{\mu\nu}+g_{\mu\nu}p_{\phi})\Big]\,,
\end{eqnarray}
where $\rho_\phi$ and $p_\phi$ are the energy density and pressure of the quintessence field, defined as
\begin{eqnarray}\label{eq10}
\rho_\phi=\frac{1}{2}g^{\mu\nu}\partial_{\mu}\phi \partial_{\nu}\phi+V(\phi)\,,\nonumber\\
p_\phi=\frac{1}{2}g^{\mu\nu}\partial_{\mu}\phi \partial_{\nu}\phi-V(\phi)\,,
\end{eqnarray}
and $u_\mu$ is the four-velocity with $u_\mu u^{\mu}=-1$. Now, we consider
a specific representation of the dissipation exponent as
\begin{eqnarray}\label{eq12}
{\Gamma(g_{\mu \nu},\phi,x^\mu)}=\int \Big(\triangledown_{\lambda}u^{\lambda}\Big)Q(\phi,x^{\alpha})\sqrt{-g}d^4x\,.
\end{eqnarray}
where $Q(\phi,x^{\alpha})$ is an arbitrary function called the \emph{dissipation function}.
In the case of the FRW geometry, in the comoving frame $u_\lambda = (1, 0, 0, 0)$
and $\triangledown_\lambda u^{\lambda}=3H$. This kind of dissipation exponent
leads to the Klein-Gordon equation of the form $\ddot{\phi}+3H(1+Q)\dot{\phi}+\frac{V(\phi)}{d\phi}=0$
which was widely studied in warm inflation models but without deriving from a variational
principle. For a dissipation exponent \eqref{eq12}, we obtain
\begin{eqnarray}
\Theta_{\mu\nu}=-\int{\Big[\triangledown_{\lambda}u^{\lambda}h_{\mu\nu}+
u^{\lambda}\frac{\partial}{\partial x^{\lambda}}h_{\mu\nu}\Big]Q(x^\alpha)\sqrt{-g}d^4x}\,,
\end{eqnarray}
where we have used the following relations
\begin{eqnarray}
\delta u^{\lambda}=-\frac{1}{2}u^{\lambda}u_{\alpha}u_{\beta}\delta
g^{\alpha\beta}\,,\hspace{0.5cm}\nonumber\\
\delta \sqrt{-g}=-\frac{1}{2}\sqrt{-g}g_{\alpha\beta}\delta g^{\alpha\beta}\,,
\end{eqnarray}
and $h_{\mu\nu}=u_{\mu}u_{\nu}+g_{\mu\nu}$ is called the projection operator.
Thus, the energy-momentum tensor of the dissipative scalar field for a
dissipation exponent \eqref{eq12} is given by
\begin{eqnarray}
T_{\mu\nu}^{(\phi)}=e^{\int \triangledown_{\lambda}u^{\lambda}Q(x^{\alpha})\sqrt{-g}d^4x}\hspace{11.8cm}\nonumber\\
\times\Bigg[\triangledown_{\mu}\phi \triangledown_{\nu}\phi-\bigg(\int{\Big[\triangledown_{\lambda}u^{\lambda}h_{\mu\nu}+u^{\lambda}\frac{\partial}{\partial x^{\lambda}}h_{\mu\nu}\Big]Q(x^\alpha)\sqrt{-g}d^4x}+g_{\mu\nu}\bigg)
\bigg(\frac{1}{2}g^{\alpha\beta}\triangledown_{\alpha}\phi \triangledown_{\beta}\phi+V(\phi)\bigg)\Bigg]
\end{eqnarray}
which depends on the metric, a vector field $u^{\lambda}$, the potential and
dissipation function. Let us assume spatial flatness, homogeneity and isotropy of the
universe and take the Friedmann-Robertson-Walker (FRW) spacetime line element as
\begin{eqnarray}\label{eq17}
ds^2=-dt^2+a^2(t)dx_i dx^i\,, \quad i=1,2,3
\end{eqnarray}
where $a(t)$ is the scale factor. So, the Lagrangian of a dissipative scalar field is taken as
\begin{eqnarray}
L_{\phi}=a^3 e^{3\int{H(t)Q(t)dt}}\bigg(\frac{1}{2}{\dot{\phi}}^2-V(\phi)\bigg)\,.
\end{eqnarray}
Then, the Klein-Gordon equation can be written as
\begin{eqnarray}\label{eq18}
\ddot{\phi}(t)+3H(t)(1+Q(t))\dot{\phi}(t)+\frac{dV(\phi)}{d\phi}=0\,.
\end{eqnarray}
This equation shows that the function $Q(t)$ acts as a novel
dissipative term in the Klein-Gordon equation. If the dissipation function
depends on the scalar field, $Q=Q(\phi(t))$, the equation of motion\eqref{eq5} leads to
\begin{eqnarray}\label{eq19}
\ddot{\phi}(t)+3H(1+Q(\phi(t)))\dot{\phi}+V'(\phi)-3\int{H(t)Q'(\phi(t))dt\bigg(\frac{1}{2}\dot{\phi}^2-V(\phi)\bigg)}=0\,,
\end{eqnarray}
where a prime denotes derivative with respect to the scalar field and a
dot marks derivative with respect to the cosmic time. The energy density
of the dissipative quintessence field in FRW metric takes the following form
\begin{eqnarray}
T^{0\,(\phi)}_0=e^{3\int{H(t)Q(t)dt}}\Big(\frac{1}{2}\dot{\phi}^2+V(\phi)\Big)=\rho_{\phi}^{(eff)}\,.
\end{eqnarray}
To calculate the effective pressure of the dissipative scalar field $p_{\phi}^{(eff)}$,
we assume the energy-momentum of the effective scalar field is covariantly conserved, that is,
\begin{eqnarray}\label{eq21}
\dot{\rho}_{\phi}^{(eff)}+3H\Big(\rho_{\phi}^{(eff)}+p_\phi^{(eff)}\Big)=0\,.
\end{eqnarray}
Hence, by using equations \eqref{eq10},\eqref{eq18} and \eqref{eq21}, we
obtain the effective pressure of the dissipative scalar field as follows
\begin{eqnarray}
p_\phi^{(eff)}=(1+Q)p_{\phi}e^{3\int{H(t)Q(t)dt}}\,.
\end{eqnarray}
Now, in the spatially flat FRW background, the $tt$ and $ii$ components of the
field equations can be written as follows
\begin{eqnarray}\label{eq22}
3H^2=\kappa^2\Big(\rho_\phi^{(eff)}+\rho_m\Big)\hspace{6cm}\nonumber\\
=\kappa^2\bigg[\Big(\frac{1}{2}\dot{\phi}^2+V(\phi)\Big)e^{3\int{H(t)Q(t)dt}}+\rho_m\bigg]\,,\hspace{2.15cm}
\end{eqnarray}
\begin{eqnarray}\label{eq23}
2\dot{H}+3H^2=-\kappa^2\big[p_{\phi}^{(eff)}+p_m]\hspace{5.56cm}\nonumber\\
=-\kappa^2\bigg[(1+Q)\Big(\frac{1}{2}\dot{\phi}^2-V(\phi)\Big)e^{3\int{H(t)Q(t)dt}}+p_m\bigg]\,.
\end{eqnarray}
By using equations \eqref{eq22} and \eqref{eq23}, we obtain the generalized
conservation equation as
\begin{eqnarray}
\frac{d}{dt}(a^3 \rho_{\phi}^{(eff)})+\frac{da^3}{dt}p_{\phi}^{(eff)}+\frac{d}{dt}(a^3\rho_m)+\frac{da^3}{dt}p_m=0\,.
\end{eqnarray}
With the previous assumption of the conservation of the effective energy density
of the dissipative quintessence field, this equation shows that the energy density
of matter is also conserved. This means that there is no matter energy transfer
between the dissipative quintessence field and ordinary baryonic matter. Hence,
the conservation of matter is also satisfied,
\begin{eqnarray}
\dot{\rho}_m+3H(\rho_m+p_m)=0\,.
\end{eqnarray}
The equation of state of the dissipative quintessence field is defined by
\begin{eqnarray}
\omega=(1+Q)\frac{p_{\phi}}{\rho_{\phi}}\,.
\end{eqnarray}
During the inflation era and in the slow-roll approximation, which the potential energy
dominates over the kinetic energy of the inflationary field, we have
$\ddot{\phi}\ll |{3H\dot{\phi}}|$ and $\dot{\phi}^2\ll V(\phi)$. So, the Friedmann
equation and equation of motion for the scalar field within the slow-roll approximation
can be written as
\begin{eqnarray}\label{eq28}	
H^2=\frac{\kappa^2}{3}V(\phi)e^{3\int{H(t)Q(t)dt}}\,,
\end{eqnarray}
\begin{eqnarray}\label{eq29}
\dot{\phi}=-\frac{V'(\phi)}{3H(1+Q)}\,.
\end{eqnarray}
We assume a universe free of ordinary baryonic matter, $\rho_m=0$. The Hubble slow-roll parameters are defined as
\begin{eqnarray}
\epsilon=-\frac{\dot{H}}{H^2}\,, \quad\quad \eta=\frac{1}{H}\frac{\ddot{H}}{\dot{H}}\,.
\end{eqnarray}
In our setup, the slow-roll parameters in terms of potential $V(\phi)$
of the scalar field can be written as
\begin{eqnarray}
\epsilon=\frac{e^{-3\int{HQdt}}}{2\kappa^2(1+Q)}\Big(\frac{V'}{V}\Big)^2\,,
\end{eqnarray}
\begin{eqnarray}
\eta={\frac {3 \Big(\frac{1}{3}\,{\kappa}^{2}{{\it V'}}^{2}{V}^{2} Q(1+Q)
{e^{3\int{HQdt}}}-{\kappa}^{4}{V}^{4}{Q}^{2} \left( 1+Q
\right) ^{2}{e^{6\int{HQdt}}}-\frac{2}{9}\,{\it V''}\,{{\it V'}}^{2}V+\frac{1}{9}\,
{{\it V'}}^{4} \Big) {e^{-3\int{HQdt}}}}{{\kappa}^{2} \left( Q{V}
^{2}{\kappa}^{2} \left( 1+Q \right) {e^{3\int{HQdt}}}-\frac{1}{3}\,{{\it V'}}^{2} \right)  \left( 1+Q \right) {V}^{2}}}\,.	
\end{eqnarray}
The slow-roll conditions for inflationary phase are $\epsilon \ll 1$ and $\eta \ll 1$
and whenever one of these parameters reaches unity, the inflation phase terminates.
We introduce the number of e-folds as
\begin{eqnarray}
N=\int_{t_{hc}}^{t_f}{H(t)dt}\,.
\end{eqnarray}
where $t_{hc}$ is the horizon crossing time and $t_f$ is the end time of inflation.
We proceed to test this model by studying the linear perturbations of the initial
fluctuations. In this regard, we study the spectrum of perturbations
produced by quantum fluctuations of the fields about their homogeneous
background values in the next section.\\

\section{Perturbations}

In this section, we explore the linear perturbation theory in dissipative quintessential inflation.
These perturbations are generated by the quantum fluctuations of both the
spacetime metric and the scalar field around the homogeneous background
solutions. The FRW metric in the spatially longitudinal gauge can
be expressed as~\cite{Bar80,Muk92,Ber95}
\begin{eqnarray}
ds^2=-[1+2\Phi(t,\textbf{x})]dt^2+a^2(t)[1-2\Psi(t,\textbf{x})]dx_i dx^i\,,
\end{eqnarray}
where $\Phi(t,\textbf{x})$ and $\Psi(t,\textbf{x})$ are the metric perturbations.
It is assumed that $\Phi$ and $\Psi$ are gauge invariant variables~\cite{Muk92}. In the
absence of these perturbations, the FRW line element is recovered. To proceed,
we consider $\phi(t,\textbf{x})=\phi(t)+\delta \phi(t,\textbf{x})$, where $\delta \phi(t, \textbf{x})$ is the
linear perturbation of the scalar field. Now, we obtain the equations of motion for
cosmological perturbations from the linearized Einstein equations. From
the $i\neq j$ components of perturbed Einstein field equations, we get $\Phi=\Psi$
which means that two metric perturbations are equal. So, we obtain the
perturbed Einstein field equations as
\begin{eqnarray}
6H\dot{\Phi}+6H^2 \Phi-2a^{-2}\triangledown^2 {\Phi}=\kappa^2
{e^{3\int{H(t)Q(t)dt}}} \bigg[\big(\Phi \dot{\phi}^2-\dot{\phi}\dot{(\delta\phi)}-V'\delta \phi \big)\nonumber \\
-\big(\frac{1}{2}\dot{\phi}^2+V \big)\int{a^3\big[-3Q(\dot{\Phi}+3H\Phi)+3H\delta Q \big]d^4 x}\bigg]\,,
\end{eqnarray}
\begin{eqnarray}
\dot{\Phi}+H\Phi=\frac{\kappa^2}{2}\dot{\phi}\,\delta \phi\, e^{e^{3\int{H(t)Q(t)dt}}}\,,
\end{eqnarray}
\begin{eqnarray}
2\ddot{\Phi}+2(2\dot{H}+3H^2)\Phi+8H\dot{\Phi}=\kappa^2e^
{3\int{H(t)Q(t)dt}}\Big\{2\Phi(1+Q)\big(\frac{1}{2}{\dot{\phi}}^2-V\big)+
a^{-2}\big(\frac{1}{2}\dot{\phi}^2-V \big)\,\nonumber\\
\times \Big(\int{5Ha^5Q(t)d^4x}+a^2\Big)\int{a^3\big[-3Q(\dot{\Phi}+
3H\Phi)+3H\delta Q \big]d^4 x}+a^{-2}\big(\frac{1}{2}\dot{\phi}^2-V \big)\nonumber\\
\times\Big[\int{(-5\dot{\Phi}-15H\Phi)a^5Q(t)d^4x}+\int{5Ha^5(\delta Q-2\Phi Q)d^4x}
-2a^2\Phi\Big]-a^{-2}\hspace{1cm}\nonumber\\
\times\Big(\int{5Ha^5Q(t)d^4x}+a^2\Big)(\Phi {\dot{\phi}}^2- \dot{\phi}\dot{(\delta\phi)}
+V' \delta{\phi})\Big\}=0\,.\hspace{4.6cm}
\end{eqnarray}
And the perturbed scalar field equation is given by
\begin{eqnarray}
\partial_i \partial^i \phi-\ddot{(\delta\phi)}+\dot{\Phi}\dot{\phi}+3\Phi \dot{\phi}
-3H(1+Q)\dot{(\delta\phi)}-\dot{\phi}\dot{(\delta\Gamma)}+2\Phi V'-V''\delta\phi=0\,,
\end{eqnarray}
where
\begin{eqnarray}
\delta \Gamma=-3\int{(\dot{\Phi}Q+3HQ\Phi-H\delta Q)a^3 d^4x}\,.
\end{eqnarray}
Scalar perturbations can be divided into the entropy (isocurvature)
perturbations which are projection orthogonal to the trajectory, and
adiabatic (curvature) perturbations which are projection parallel to
the trajectory. For the single scalar field scenario, the perturbations
are adiabatic perturbations. Now, we introduce a gauge-invariant
primordial curvature perturbation $\zeta$, on scales outside the horizon as~\cite{Bar83}
\begin{eqnarray}\label{eq36}
\zeta=\Psi-\frac{H}{\dot{\rho}_{\phi}}\delta \rho_{\phi}
\end{eqnarray}
For uniform density hypersurfaces, $\delta \rho_{\phi}=0$, this equation
leads to the curvature perturbation,$\Psi$. The time evolution
of equation\eqref{eq36} is given as~\cite{Wan00}
\begin{eqnarray}\label{eq37}
\dot{\zeta}=\frac{H}{\rho_{\phi}+p_{\phi}}\delta p_{nad}\,,
\end{eqnarray}
which shows that the change in the curvature perturbation
with uniform density hypersurfaces, on large scales, is caused
by the non-adiabatic part of the pressure perturbation. We note
that this result is independent of the gravitational field equations.
If the pressure perturbation is adiabatic on large scales, then $\zeta$ is a
constant. The pressure perturbation (in any gauge) can be divided
into two parts, adiabatic and entropic (non-adiabatic)~\cite{Wan00}
\begin{eqnarray}\label{eq38}
\delta p=c_s^2 \delta \rho_{\phi}+\dot{p}_{\phi}\Upsilon\,,
\end{eqnarray}
where $c_s^2=\frac{\dot{p}_{\phi}}{\dot{\rho}_{\phi}}$ is the sound effective
velocity. The non-adiabatic part of the pressure perturbation is
defined $\delta p_{nad}=\dot{p}_{\phi}\Upsilon$. $\Upsilon$ is called
entropy perturbation which is gauge invariant and expresses the
displacement between hypersurfaces of uniform pressure and uniform density
 \begin{eqnarray}
 \Upsilon\equiv \frac{\delta p_{\phi}}{\dot{p}_{\phi}}+\frac{\delta \rho_{\phi}}{\dot{\rho}_{\phi}}\,.
 \end{eqnarray}
By using equation~\eqref{eq38} and within the slow-roll conditions, we find
\begin{eqnarray}
\delta p_{nad}=0\,.
\end{eqnarray}
This means that the non-adiabatic part of the pressure perturbation is zero,
so the pressure perturbation is adiabatic. Hence, from equation~\eqref{eq37}, we obtain
\begin{eqnarray}
\dot{\zeta}=0\,.
\end{eqnarray}
The curvature perturbation on uniform density hypersurfaces, in terms
of scalar field fluctuations on spatially flat hypersurfaces is expressed as~\cite{Lyt09}
\begin{eqnarray}\label{eq46}
\zeta=-\frac{H\delta \phi}{\dot{\phi}}\,.
\end{eqnarray}
Moreover, the field fluctuations in the Hubble crossing and in the
slow-roll condition are expressed as
\begin{eqnarray}
\langle{(\delta \phi)^2}\rangle=\frac{H^2}{4\pi^2}\,.
\end{eqnarray}
This relation is independent of the underlying gravity theory for the
massless field in de Sitter space. Eventually, the density perturbations
in adiabatic perturbations are given by the following form~\cite{Lid93}
\begin{eqnarray}\label{eq48}
A_s^2=\frac{\langle\zeta^2\rangle}{V(\phi)}\,.
\end{eqnarray}
By using Eqs.~\eqref{eq46}-\eqref{eq48}, we obtain
\begin{eqnarray}
A_s^2=\frac{H^4}{4\pi^2 {\dot{\phi}}^2}\,.
\end{eqnarray}
The density perturbations in our dissipative quintessence model can be derived as
\begin{eqnarray}
A_s^2=\frac{\kappa^2(1+Q)^2}{12\pi^2}\bigg(\frac{V}{V'}\bigg)^2e^{9\int{H(t)Q(t)dt}}\,.
\end{eqnarray}
The scalar spectral index is defined as
\begin{eqnarray}
n_s-1=\frac{d\ln A_s^2}{d\ln k}
\end{eqnarray}
The relation between wave number and e-folds number is expressed as
\begin{eqnarray}
d\ln k(\phi)=dN(\phi)\,.
\end{eqnarray}
In our dissipative quintessence setup and within the slow-roll approximation, the scalar spectral
index takes the following analytical form
\begin{eqnarray}
n_s-1=\frac{2\dot{Q}}{(1+Q)H}-\frac{2e^{-3\int{H(t)Q(t)dt}}}
{\kappa^2(1+Q)}\bigg(\frac{{V'}^2}{V^2}-\frac{V''}{V}\bigg)+9Q\,.
\end{eqnarray}
The tensor perturbations amplitude of a given state when leaving the Hubble radius are
\begin{eqnarray}
A_T^2=\frac{4\kappa^2}{25\pi}H^2\bigg|_{k=aH}
\end{eqnarray}
In our model, we find
\begin{eqnarray}
A_T^2=\frac{4\kappa^4}{75\pi}Ve^{3\int{H(t)Q(t)dt}}\,.
\end{eqnarray}
The tensor spectral index is defined as
\begin{eqnarray}
n_T=\frac{d\ln A_T^2}{d\ln k}\,.
\end{eqnarray}
So, we obtain the tensor spectral index in terms of the potential
$V(\phi)$ of the scalar field and the dissipation function as follows
\begin{eqnarray}
n_T=\frac{16\pi e^{-6\int{H(t)Q(t)dt}}}{25(1+Q)^2\kappa^2}\bigg(\frac{{V'}^2}{V}\bigg)\,,
\end{eqnarray}
which depends on the potential and dissipation function.
Another important parameter that provides information about
the perturbations is the tensor-to-scalar ratio defined as
\begin{eqnarray}
r=\frac{A_T^2}{A_s^2}\,,
\end{eqnarray}
where $A_T^2$ and $A_s^2$ are given by Eqs. (55) and (50) respectively.

In the next section, we perform numerical analysis on the parameters space of
our dissipative quintessential inflation model. In this
regard, we consider different forms of the dissipation function and potential.
We confront our results with the latest observational data to constraint our model.

\section{Observational Constraints}

In this section, we examine viability of our dissipative
quintessential inflation in confrontation with observational data.
We firstly calculate numerical values of the main observables in our setup and
then compare these calculated numerical results with the latest observational data
from Planck and BICEP/Keck joint data sets. We perform our numerical analysis mainly on the
perturbation parameters $n_{s}$ and $r$. In this regard, we need to
adopt some specific functions for $V(\phi)$ and $Q$. After choosing
the mentioned functions appropriately, we obtain the scalar spectral index and
tensor-to-scalar ratio in terms of the model's parameters that
prepares us to perform a numerical analysis on the parameters space
and comparing the obtained results with observation. We
compare the numerical results with both Planck2018 TT, TE,
EE+lowE+lensing+BAO+BK\textbf{14} joint data~\cite{pl18a,pl18b} and Planck2018 TT,
TE, EE+lowE+lensing+BAO +BK\textbf{18} joint data~\cite{Bi18a,Bi18b}. In what follows,
we consider two cases for $Q$. First, the case where $Q$
is a constant. Second, the case where $Q$ is a function of the scalar
field as $Q=\alpha \phi^{n}$ (with $\alpha$ and $n$ being constant
parameters).\\
We note that as our strategy in all of our forthcoming numerical analysis, we have adopted
two values of the e-folds numbers; $N=55$ and $N=60$ and then we compare the results through figures and tables. For exponential potential
we re-scale the value of the parameter $\beta$ (to be defined later) as $\beta=1$. For every sample values of the parameter
$\lambda$ (to be defined later), we consider observational values of both $n_s$ and $r$ at $68\%$ and $95\%$
confidence levels. From those constraints at both confidence levels, we solve the system
of equations corresponding to $n_s$ and $r$ and obtain the constraints as presented in the
forthcoming Tables. It is worth mentioning that there are some values of the parameters leading to
observational constraint on $n_s$ but not on $r$ (and vice versa). We drop them away and just
consider the ranges that make both of $n_s$ and $r$ observationally viable simultaneously.

\subsection{ $Q=constant$}

The first case we study here is the one where the dissipative function
is a constant parameter as $Q=Q_{0}$. Now, from equations\eqref{eq28}
and\eqref{eq29}, we obtain the scale factor in this case as
\begin{eqnarray}
a(\phi)=\bigg[3Q_0(1+Q_0)\kappa^2\int{\frac{V}{V'}d\phi}\bigg]^{-\frac{1}{3Q_0}}\,.
\end{eqnarray}
We also define the number of e-folds during inflation in terms of the
scale factor as
\begin{eqnarray}
N=\int{Hdt}=\int_{\phi_{hc}}^{\phi_{f}}{\frac{a'(\phi)}{a(\phi)}d\phi}\,.
\end{eqnarray}
In the constant dissipative parameter case, the scalar spectral
index and the tensor-to-scalar ratio take the following forms respectively
\begin{eqnarray}\label{eq61}
n_s=1-2\,\frac{e^{-3Q_0\int{H dt}}}{\kappa^2(1+Q_0)}\bigg(\frac{{V'}^2}{V^2}-\frac{{V''}}{V}\bigg)+9Q_0\,,
\end{eqnarray}
\begin{eqnarray}\label{eq62}
r=\frac{16\pi{e^{-6Q_0\int{H dt}}}}{25\kappa^2(1+Q_0)^2}\bigg(\frac{{V'}^2}{V}\bigg)\,.
\end{eqnarray}
which are functions of the scalar field potential. Therefore, to complete our
analysis and discussions, we have to chose some specific forms of the scalar field potential.
We consider two types of potentials as exponential and power-law potential.

\subsubsection{Exponential Potential}

The exponential potential that we consider here, is defined
as~\cite{Gr18}
\begin{eqnarray}\label{eq63}
V(\phi)= M^4e^{-\lambda \phi}\,,
\end{eqnarray}
leading to the following expression for the number of e-folds parameter
\begin{eqnarray}\label{eq64}
N=-\frac{1}{3Q_0}\,(\ln \phi_f-\ln \phi_{hc})\,.
\end{eqnarray}
Using the condition $\epsilon=1$ at the end of inflation phase, we find $\phi_{hc}$ from
equation \eqref{eq64} for a given $N$ and then substitute the result into equations \eqref{eq61}
and \eqref{eq62} for the exponential potential. In this way, we find the
scalar spectral index and the tensor-to-scalar ratio in terms of $N$
as follows
\begin{eqnarray}
n_s=1+9Q_0\,,
\end{eqnarray}
\begin{eqnarray}
r=\frac{16\pi M^{4}\lambda^{2} e^{-6Q_{0}N} e^{\frac{e^{-3Q_{0}N}(2+3Q_{0})}{3Q_{0}}}}{25\kappa^{2}(1+Q_{0})^{2}}
\end{eqnarray}
Now, we study these parameters numerically to find some
observational constraints on the model's parameters. By performing
the numerical analysis, we obtained the ranges of the parameters
$\lambda$ and $Q_0$ for $N=55$ and $N=60$ as shown in figure 1 and 2 respectively. The $r-n_{s}$ plane in
these cases and in the background of both Planck2018 TT, TE,
EE+lowE+lensing+BAO+BK\textbf{14} and Planck2018 TT, TE,
EE+lowE+lensing+BAO +BK\textbf{18} data are shown in figure 3 and 4 for $N=55$ and $N=60$ respectively. By
analyzing these results, we have found some constraints on the
model's parameter space that are summarized in tables 1 and 2 for $N=55$ and $N=60$ respectively. As these tables show, the dissipation
factor is negative and so small in measure. Another important point to note is related to the
case that there is no dissipation. As figures 1 and 2 show, the vertical red strips are the acceptable ranges of parameter $Q_{0}$ for
observational consistency of $n_{s}$ with $N=55$ and $N=60$ respectively. Obviously the case $Q_{0}=0$, that is,
no dissipation, is not observationally acceptable in this framework.
Therefore, it is concluded that the dissipative quintessential inflation is more realistic and reliable from observational ground than the
standard non-dissipative quintessential inflation with the same adopted potentials. Also a comparison between figures 3 and 4 with $N=55$ and $N=60$ respectively,
shows that our dissipative quintessential inflation model with $N=55$ is observationally more viable than the case with $N=60$.

\begin{figure}
\begin{center}{\includegraphics[width=.66 \textwidth,origin=c,angle=0]{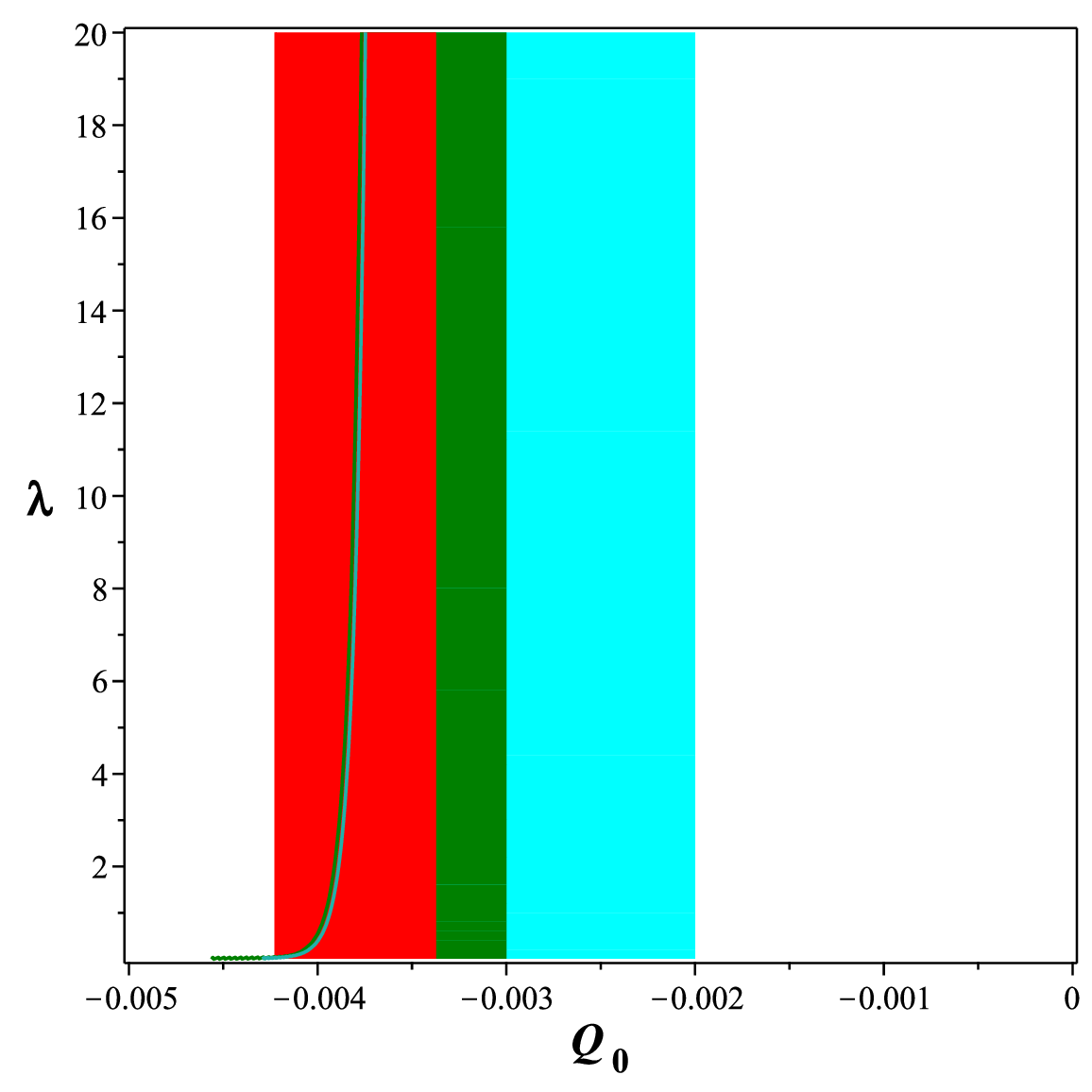}}
\end{center}
\caption{\label{fig1}\small {Ranges of the parameters $\lambda$ and
$Q_{0}$ in our model (with constant $Q_{0}$ and exponential potential)
leading to observationally viable values of $r$ and $n_{s}$ with $\textbf{N=55}$. The red
region shows the values of the parameters ($Q_{0}$,$\lambda$) leading to observationally
viable values of $n_{s}$ in confrontation with Planck2018 TT, TE,
and EE+lowE+lensing+BAO+BK\textbf{14} data. The cyan region shows the values
of the parameters ($Q_{0}$,$\lambda$) leading to observationally viable values of $r$ in
confrontation with Planck2018 TT, TE, and EE+lowE+lensing+BAO+BK\textbf{14}
data. The green region shows the values of the parameters ($Q_{0}$,$\lambda$) leading to
observationally viable values of $r$ in confrontation with
Planck2018 TT, TE, and EE+lowE+lensing+BAO+BK\textbf{18} data.}}
\end{figure}

\begin{figure}
\begin{center}{\includegraphics[width=.66 \textwidth,origin=c,angle=0]{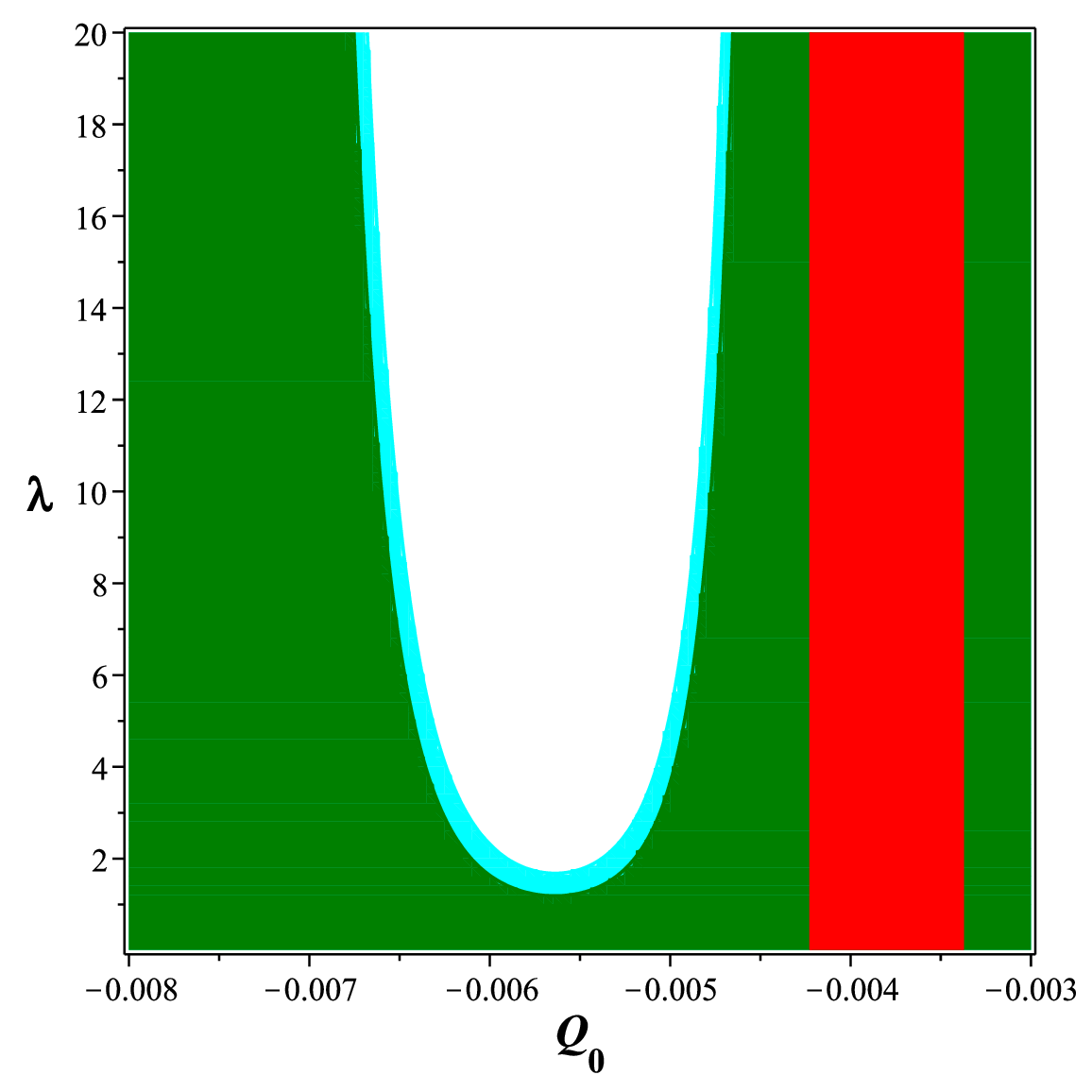}}
\end{center}
\caption{\label{fig1}\small {Ranges of the parameters $\lambda$ and
$Q_{0}$ in our model (with constant $Q_{0}$ and exponential potential)
leading to observationally viable values of $r$ and $n_{s}$ with $\textbf{N=60}$. The red
region shows the values of the parameters ($Q_{0}$,$\lambda$) leading to observationally
viable values of $n_{s}$ in confrontation with Planck2018 TT, TE,
and EE+lowE+lensing+BAO+BK\textbf{14} data. The cyan region shows the values
of the parameters ($Q_{0}$,$\lambda$) leading to observationally viable values of $r$ in
confrontation with Planck2018 TT, TE, and EE+lowE+lensing+BAO+BK\textbf{14}
data. The green region shows the values of the parameters ($Q_{0}$,$\lambda$) leading to
observationally viable values of $r$ in confrontation with
Planck2018 TT, TE, and EE+lowE+lensing+BAO+BK\textbf{18} data.}}
\end{figure}

\begin{figure}
\begin{center}{\includegraphics[width=.66 \textwidth,origin=c,angle=0]{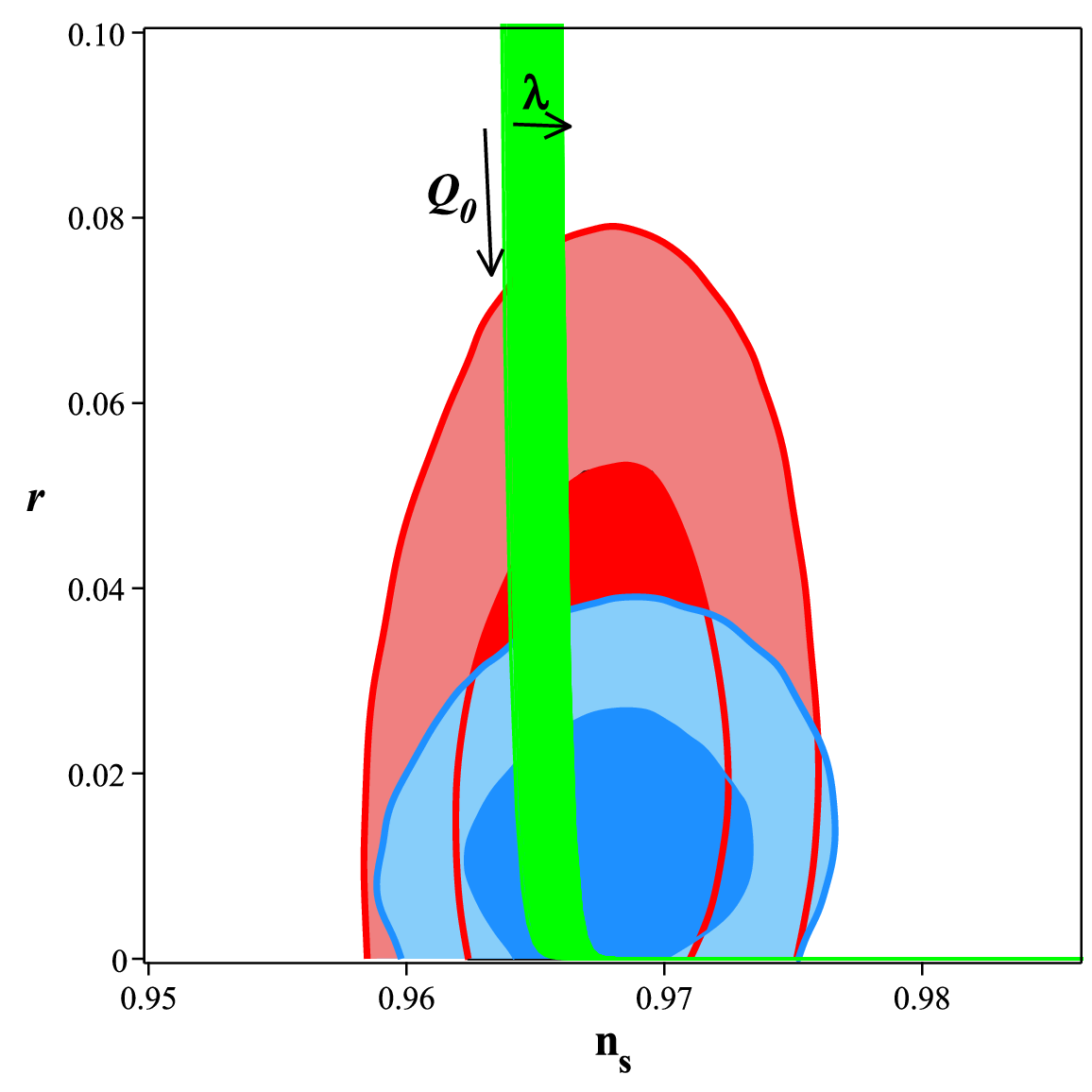}}
\end{center}
\caption{\label{fig2}\small {$r-n_{s}$ plane of our model
with $\textbf{N=55}$, constant $Q_{0}$ and exponential potential (green region) in the background of
Planck2018 TT, TE, and EE+lowE+lensing+BAO+BK\textbf{14} joint data (red region) and Planck2018
TT, TE, and EE+lowE+lensing+BAO+BK\textbf{18} joint data (blue region). The ranges of $\lambda$ and $Q_{0}$ are as $1<\lambda<20$ and $-0.00400<Q_{0}<-0.00200$.}}
\end{figure}

\begin{figure}
\begin{center}{\includegraphics[width=.66 \textwidth,origin=c,angle=0]{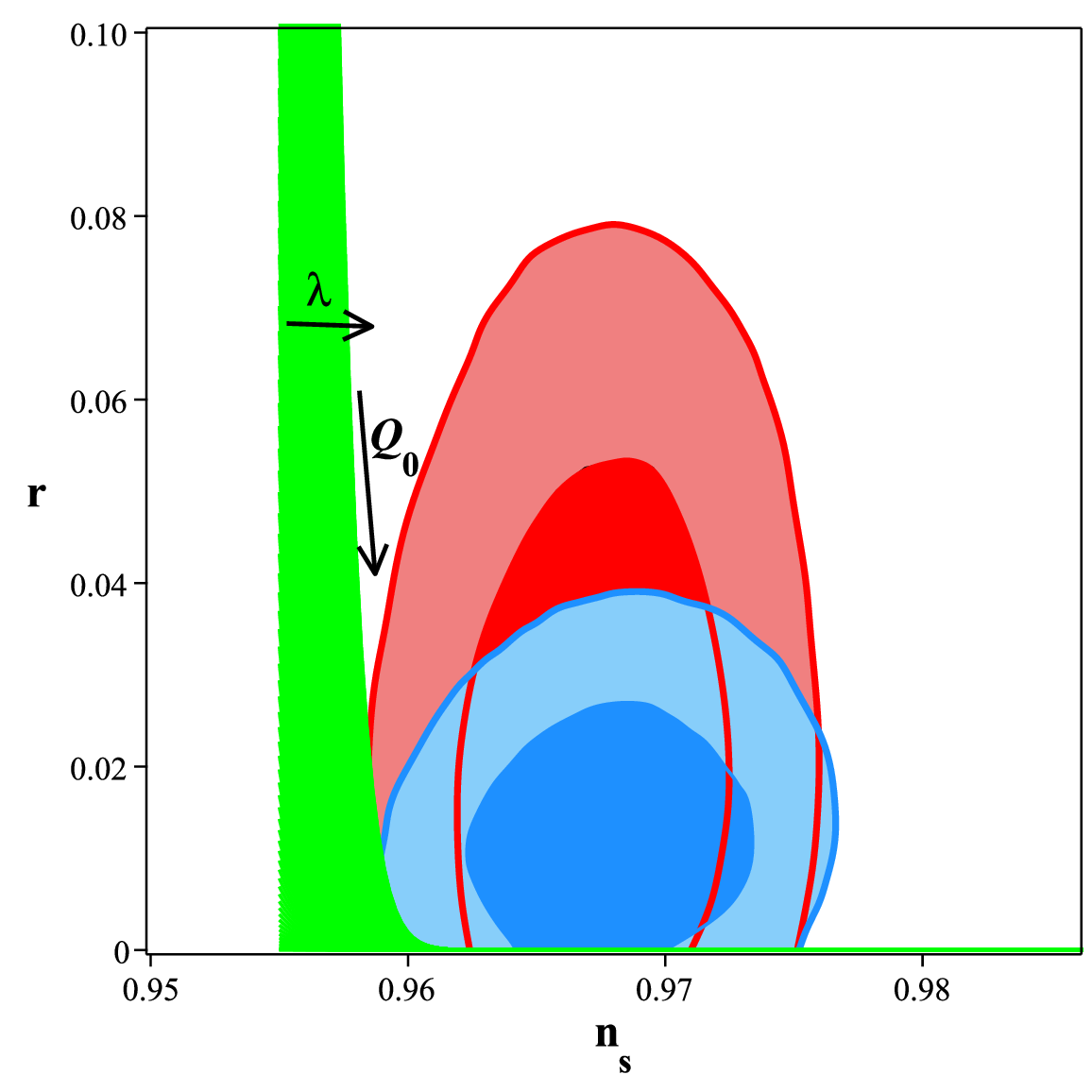}}
\end{center}
\caption{\label{fig2}\small {$r-n_{s}$ plane of our model
with $\textbf{N=60}$, constant $Q_{0}$ and exponential potential (green region) in the background of
Planck2018 TT, TE, and EE+lowE+lensing+BAO+BK\textbf{14} joint data (red region) and Planck2018
TT, TE, and EE+lowE+lensing+BAO+BK\textbf{18} joint data (blue region). The ranges of $\lambda$ and $Q_{0}$ are as $1<\lambda<20$ and $-0.00400<Q_{0}<-0.00200$.}}
\end{figure}

\begin{table*} \tiny\tiny\caption{\small{\label{tab1} Ranges
of the parameter $Q_{0}$, for some sample values of $\lambda$ in
which the tensor-to-scalar ratio and scalar spectral index of our
model (with constant $Q_{0}$ and exponential potential), are consistent
with different datasets with $\textbf{N=55}$.}}
\begin{center}
\tabcolsep=0.05cm\begin{tabular}{cccccc}
\\ \hline \hline \\ & Planck2018 TT,TE,EE+lowE & Planck2018 TT,TE,EE+lowE&
Planck2018 TT,TE,EE+lowE&Planck2018 TT,TE,EE+lowE
\\
& +lensing+BK\textbf{14}+BAO &
+lensing+BK\textbf{14}+BAO&lensing+BK\textbf{18}+BAO&lensing+BK\textbf{18}+BAO
\\
\hline \\$\lambda$& $68\%$ CL & $95\%$ CL &$68\%$ CL & $95\%$ CL
\\
\hline\hline \\  $4$ & \quad$-0.00385<Q_0<-0.00320$\quad
&\quad $-0.00387<Q_0<-0.00275$\quad
&\quad $-0.00383<Q_0<-0.00335$ \quad & \quad$-0.00385<Q_0<-0.00275$\quad \\ \\
\hline
\\$10$& \quad$-0.00380<Q_0<-0.00320$ \quad&\quad $-0.00382<Q_0<-0.00276$\quad
&\quad$-0.00378<Q_0<-0.00330$\quad&\quad $-0.00379<Q_0<-0.00276$\quad
\\ \\ \hline\\
$16$& \quad$-0.00377<Q_0<-0.00320$ \quad&\quad$-0.00382<Q_0<-0.00277$ \quad&
\quad$-0.00375<Q_0<-0.00327$\quad&\quad $-0.00376<Q_0<-0.00277$\quad \\ \\
\hline \hline
\end{tabular}
\end{center}
\end{table*}

\begin{table*} \tiny\tiny\caption{\small{\label{tab1} Ranges
of the parameter $Q_{0}$, for some sample values of $\lambda$ in
which the tensor-to-scalar ratio and scalar spectral index of our
model (with constant $Q_{0}$ and exponential potential), are consistent
with different datasets with $\textbf{N=60}$.}}
\begin{center}
\tabcolsep=0.05cm\begin{tabular}{cccccc}
\\ \hline \hline \\ & Planck2018 TT,TE,EE+lowE & Planck2018 TT,TE,EE+lowE&
Planck2018 TT,TE,EE+lowE&Planck2018 TT,TE,EE+lowE
\\
& +lensing+BK\textbf{14}+BAO &
+lensing+BK\textbf{14}+BAO&lensing+BK\textbf{18}+BAO&lensing+BK\textbf{18}+BAO
\\
\hline \\$\lambda$& $68\%$ CL & $95\%$ CL &$68\%$ CL & $95\%$ CL
\\
\hline\hline \\  $4$ & \quad$-0.00419<Q_0<-0.00320$\quad
&\quad $-0.00462<Q_0<-0.00275$\quad
&\quad $-0.00398<Q_0<-0.00332$ \quad & \quad$-0.00448<Q_0<-0.00275$\quad \\ \\
\hline
\\$10$& \quad$-0.00420<Q_0<-0.00321$ \quad&\quad $-0.00463<Q_0<-0.00276$\quad
&\quad$-0.00399<Q_0<-0.00333$\quad&\quad $-0.00449<Q_0<-0.00276$\quad
\\ \\ \hline\\
$16$& \quad$-0.00421<Q_0<-0.00322$ \quad&\quad$-0.00462<Q_0<-0.00277$ \quad&
\quad$-0.00401<Q_0<-0.00322$\quad&\quad $-0.00451<Q_0<-0.00277$\quad \\ \\
\hline \hline
\end{tabular}
\end{center}
\end{table*}

As another comparison, in Ref.~\cite{Santos2023} the authors studied warm inflation with a
$\beta$-exponential potential. There the authors considered a dissipation coefficient
$\Gamma$ explicitly dependent on the temperature, $T$, and then investigated the consequences
of this setup in the inflationary dynamics. In our case however, we consider the dissipation
coefficient as a function of the quintessence field, $\phi$, as the inflaton. Nevertheless,
a common feature of these two approaches is the possibility of realizing cosmic inflation
in agreement with current CMB data in both weak and strong dissipation regimes. Overall,
the results obtained in the present study are in agreement with the results reported in the mentioned reference,~\cite{Santos2023}.

\subsubsection{Power-Law Potential}

In the case of constant $Q=Q_{0}$, the second potential we consider is the power-law potential
given by
\begin{equation}\label{eq67}
V(\phi)=\beta \phi^b\,,
\end{equation}
where $b$ and $\beta$ are constant parameters. We note that just as an example, for the case
of $V(\phi)=\frac{1}{2}m^{2}\phi^{2}$, the value of the parameter $m$ as the inflaton mass is
constraint from CMB measurement to be $m=1.4 \times 10^{13}$GeV. Then, we obtain the number
of e-folds during the inflation with the power-law potential as
follows
\begin{eqnarray}\label{eq68}
N=-\frac{2}{3Q_0}\,(\ln \phi_f-\ln \phi_{hc})\,.
\end{eqnarray}
By assuming $\phi_{hc}\gg \phi_{f}$, we find $\phi_{hc}$ from
equation \eqref{eq68}. Therefore, the scalar spectral index and the
tensor-to-scalar ratio in terms of $N$ are given by
\begin{eqnarray}
n_s={\frac {9\,{\kappa}^{2} \left( Q_0+{\frac{1}{9}} \right) \left(
1+Q_0 \right) -2b\,{{\rm e}^{-6\,Q_0N}}}{{\kappa}^{2} \left( 1+Q_0
\right) }}\,,
\end{eqnarray}
and
\begin{eqnarray}
r={\frac {16\,\pi\,\beta\,{b}^{2}{{\rm e}^{{\frac{3}{2}}\, \left(
b-6\right)Q_0N} }}{25\,{\kappa}^{2} \left( 1+Q_0 \right) ^{2}}}\,.
\end{eqnarray}
We have performed a numerical analysis and found that there is no
consistency between our model in this case and the observational
data.
\subsection{$Q=\alpha\phi^n$}

Now, we consider the dissipation parameter to be a function of
the scalar field as $Q=\alpha\phi^{n}$, where $n$ and $\alpha$ are
constant parameters. In this case, the Klein Gordon equation
\eqref{eq19} within the slow-roll approximation is expressed as
\begin{eqnarray}
\dot{\phi}=-\frac{V'+3V\int{H(t)Q'(\phi(t))dt}}{3H(1+Q(\phi(t)))}\,.
\end{eqnarray}
We find the following expressions for the spectral index and
tensor-to-scalar ratio, in terms of the potential $V(\phi)$ and the
dissipation function $Q(\phi)$
\begin{eqnarray}\label{eq72}
n_s=1-\frac{2Q'}{\kappa^2(1+Q)^2}\bigg(\frac{V'+3V\int{HQ'dt}}{V}\bigg)-
\frac{2e^{-3\int{HQdt}}}{\kappa^2(1+Q)}\bigg(\frac{V'(V'+3V\int{HQ'dt})}{V^2}\bigg)\nonumber\\
-\frac{2e^{-3\int{HQdt}}}{\kappa^2(1+Q)}\bigg(\frac{V''+3V'\int{HQ'dt}}{V}\bigg)-\frac{6Q'V}{V'+3V\int{HQ'dt}}+9Q\,,\hspace{2.2cm}
\end{eqnarray}
\begin{eqnarray}\label{eq73}
r=\frac{16\pi
	e^{-6\int{HQdt}}}{25\kappa^2(1+Q)^2}\bigg(\frac{\big(V'+3V\int{HQ'dt}\big)^2}{V}\bigg)\,.
\end{eqnarray}
In what follows, we consider exponential and power-law potentials
to study the viability of our model with $Q=\alpha\phi^{n}$.

\subsubsection{Exponential Potential}

With the exponential potential \eqref{eq63}, we obtain the amplitude of the density
and tensor perturbations as follows
\begin{eqnarray}
A_s^2={\frac {{\kappa}^{6} \lambda^2\left( 1+\alpha\,{\phi}^{n}
\right) ^{2}}{27{\pi}^{2} \left({\alpha}^{2}{
\kappa}^{2}{\phi}^{2\,n}+2\,\alpha{\kappa}^{2}\,{\phi}^{n}-\frac{2\lambda^2}{3}
\right) ^{2}}{{\rm exp}\bigg\{{\frac {9\,\alpha{\kappa}^{2} \left(
\alpha\, \left( n+1 \right) {\phi}^{2\,n+1}+ \left( 2\,n+1 \right)
{\phi}^{n+1} \right) }{\lambda\, \left( n+1 \right)  \left( 2\,n+1
\right) }}\bigg\}}}\,,
\end{eqnarray}
\begin{eqnarray}
A_T^2={\frac {4\,{\kappa}^{4}{M}^{4}}{75\,\pi}{{\rm exp}\bigg\{{{\frac {3\,{\alpha}^{2}{\kappa}^{2} \left( n+1 \right)
{\phi}^{2\,n+1} -2\,  \left( n+\frac{1}{2} \right) \left(
-3\,\alpha\,{\kappa}^{2}{\phi}^{n+1}+{\lambda}^{2} \left( n+1
\right)\phi  \right) }{\lambda\, \left( n+1\right)  \left( 2\,n+1
\right) }}}}\bigg\}}\,.\hspace{0.9cm}
\end{eqnarray}
By using equations \eqref{eq72} and \eqref{eq73}, we obtain the
spectral index and tensor-to-scalar ratio in terms of the model
parameters, which we do not present here due to the lengthy expressions.
By performing a numerical analysis, we have found that
in this case with both $N=55$ and $N=60$ the model is consistent with observational data, at
least in some subspaces of the model's parameter space. In figures 5 and 6 we
have shown the observationally viable ranges of the parameters $\lambda$ and
$n$, in confrontation with both Planck2018 TT, TE, and
EE+lowE+lensing+BAO+BK\textbf{14} and Planck2018 TT, TE, and
EE+lowE+lensing+BAO+BK\textbf{18} data for e-folds numbers $N=55$ and $N=60$, respectively.
The $r-n_{s}$ plane for these cases in the background of Planck2018 TT, TE, and
EE+lowE+lensing+BAO+BK\textbf{14} and Planck2018 TT, TE, and
EE+lowE+lensing+BAO+BK\textbf{18} data are shown in figures 7 and 8 for e-folds numbers $N=55$ and $N=60$, respectively. From our
numerical analysis, there are some constraints on the parameter $n$,
which for some values of $\lambda$ are summarized in tables 3 and 4 for e-folds numbers $N=55$ and $N=60$, respectively.

\begin{figure}
\begin{center}{\includegraphics[width=.66 \textwidth,origin=c,angle=0]{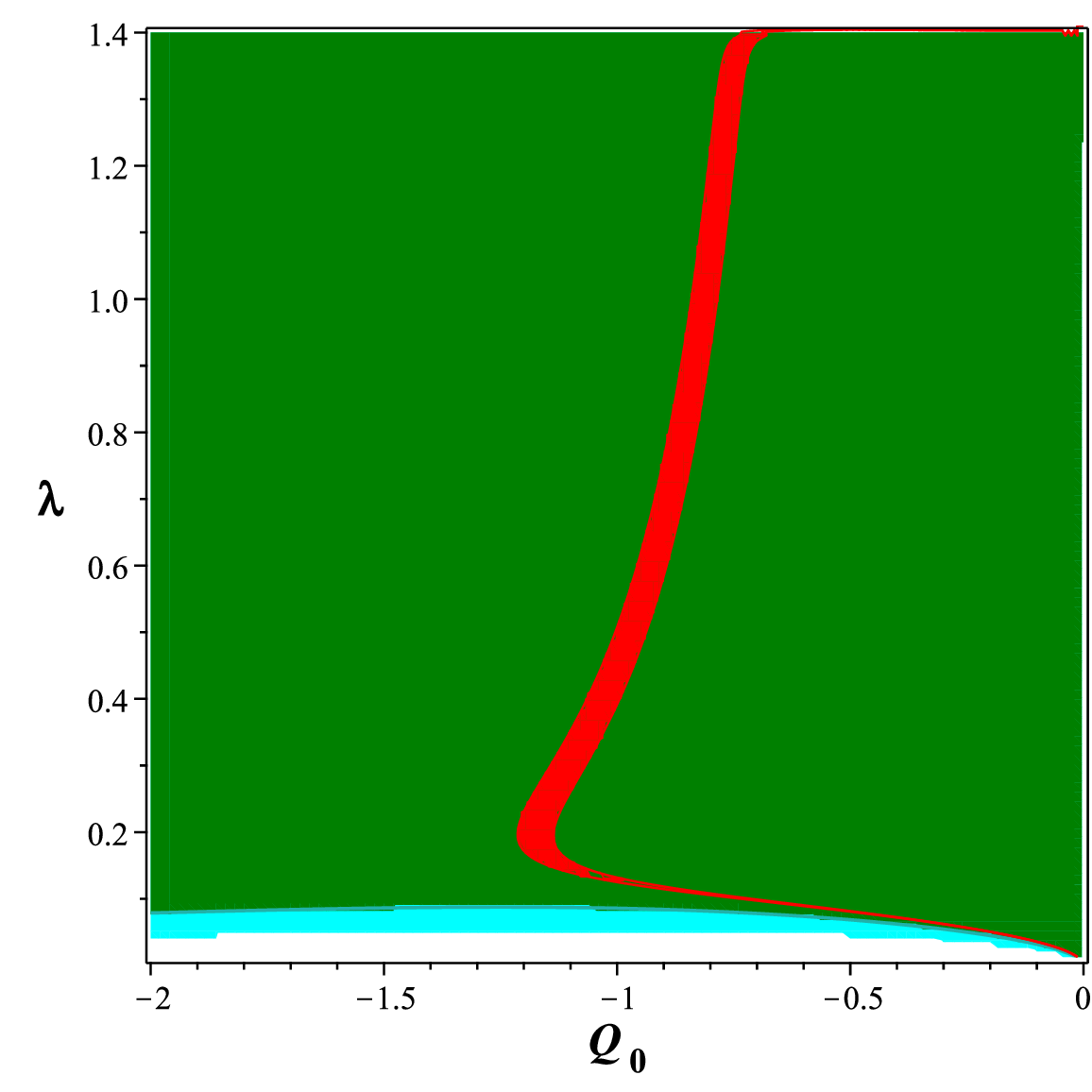}}
\end{center}
\caption{\label{fig3}\small {Ranges of the parameters $\lambda$ and
$n$ in our model with $Q=\alpha\phi^{n}$ and exponential potential,
leading to observationally viable values of $r$ and $n_{s}$ with e-folds number $\textbf{N=55}$. The red
region shows the values of the parameters leading to observationally
viable values of $n_{s}$ in confrontation with Planck2018 TT, TE,
and EE+lowE+lensing+BAO+BK\textbf{14} data. The cyan region shows the values
of the parameters leading to observationally viable values of $r$ in
confrontation with Planck2018 TT, TE, and EE+lowE+lensing+BAO+BK\textbf{14}
data. The green region shows the values of the parameters leading to
observationally viable values of $r$ in confrontation with
Planck2018 TT, TE, and EE+lowE+lensing+BAO+BK\textbf{18} data.}}
\end{figure}

\begin{figure}
\begin{center}{\includegraphics[width=.66 \textwidth,origin=c,angle=0]{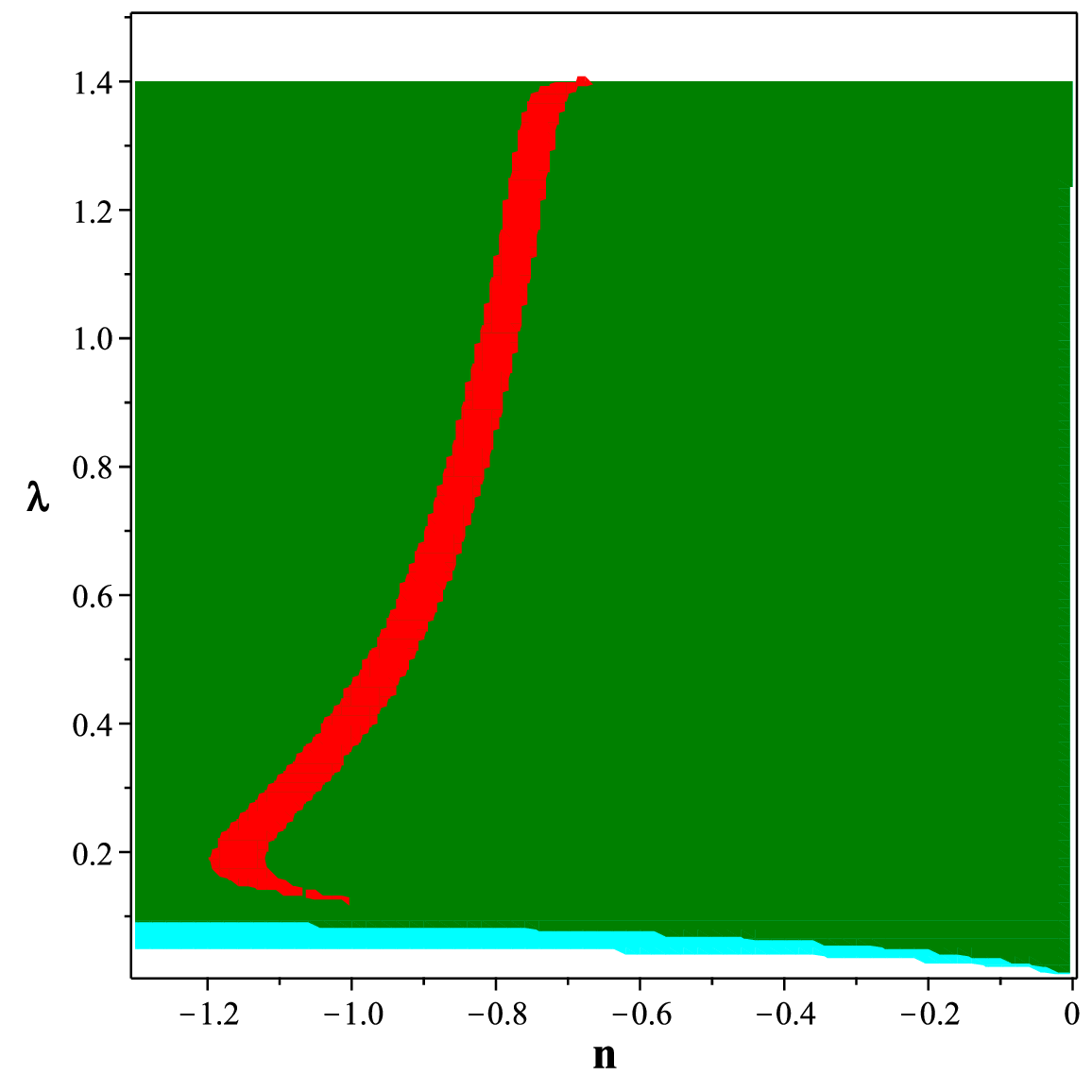}}
\end{center}
\caption{\label{fig3}\small {Ranges of the parameters $\lambda$ and
$n$ in our model with $Q=\alpha\phi^{n}$ and exponential potential,
leading to observationally viable values of $r$ and $n_{s}$ with e-folds number $\textbf{N=60}$. The red
region shows the values of the parameters leading to observationally
viable values of $n_{s}$ in confrontation with Planck2018 TT, TE,
and EE+lowE+lensing+BAO+BK\textbf{14} data. The cyan region shows the values
of the parameters leading to observationally viable values of $r$ in
confrontation with Planck2018 TT, TE, and EE+lowE+lensing+BAO+BK\textbf{14}
data. The green region shows the values of the parameters leading to
observationally viable values of $r$ in confrontation with
Planck2018 TT, TE, and EE+lowE+lensing+BAO+BK\textbf{18} data.}}
\end{figure}

\begin{figure}
\begin{center}{\includegraphics[width=.66 \textwidth,origin=c,angle=0]{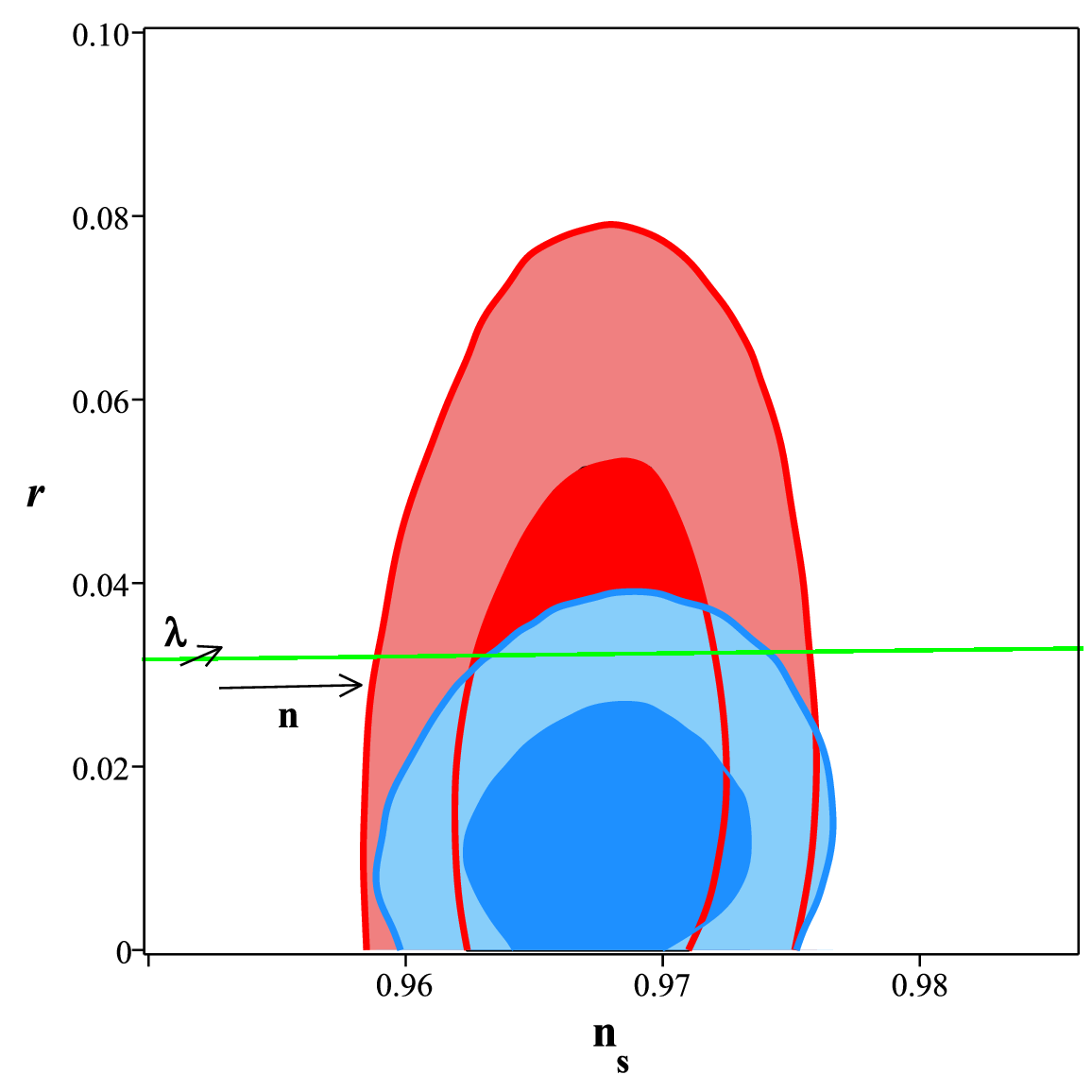}}
\end{center}
\caption{\label{fig4}\small {The $r-n_{s}$ plane with $Q=\alpha\phi^{n}$
and exponential potential (green region) in the background of Planck2018 TT, TE, and EE+lowE+lensing+BAO+BK\textbf{14} joint data (red region) and
Planck2018 TT, TE, and EE+lowE+lensing+BAO+BK\textbf{18} joint data (blue region) with $\textbf{N=55}$. The ranges of $\lambda$ and $n$ are as $0.1<\lambda<1.5$ and $-1.150<n<-0.70$.}}
\end{figure}

\begin{figure}
\begin{center}{\includegraphics[width=.66 \textwidth,origin=c,angle=0]{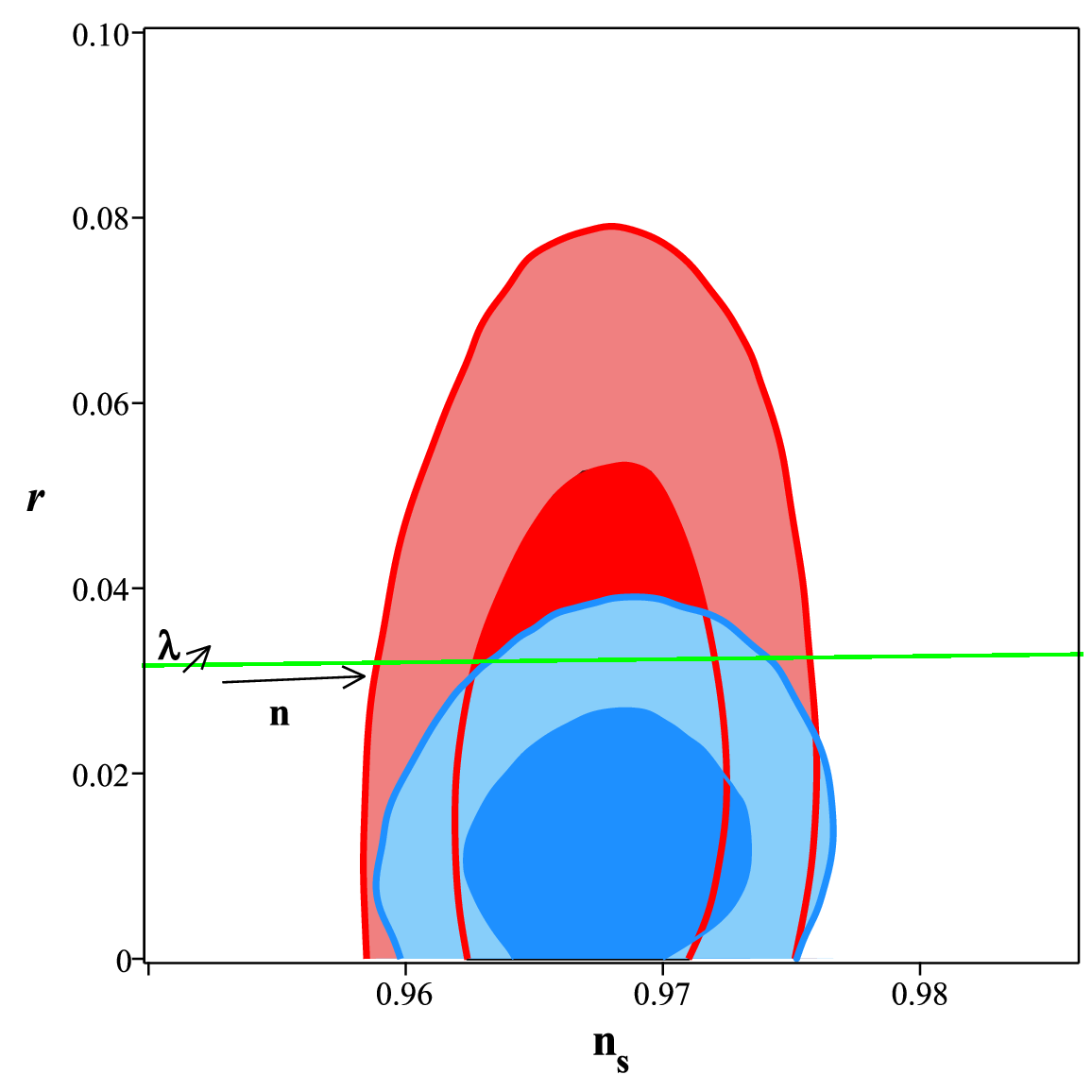}}
\end{center}
\caption{\label{fig4}\small {The $r-n_{s}$ plane with $Q=\alpha\phi^{n}$
and exponential potential (green region) in the background of Planck2018 TT, TE, and EE+lowE+lensing+BAO+BK\textbf{14} joint data (red region) and
Planck2018 TT, TE, and EE+lowE+lensing+BAO+BK\textbf{18} joint data (blue region)with $\textbf{N=60}$. The ranges of $\lambda$ and $n$ are as $0.1<\lambda<1.5$ and $-1.150<n<-0.70$.}}
\end{figure}

\begin{table*} \tiny\tiny\caption{\small{\label{tab2} Ranges
of the parameter $n$ for some sample values of $\lambda$ in which
the tensor-to-scalar ratio and scalar spectral index of our model
with $Q=\alpha\phi^{n}$ and exponential potential are consistent
with different data sets with $\textbf{N=55}$.}}
\begin{center}
\tabcolsep=0.05cm\begin{tabular}{cccccc}
\\ \hline \hline \\ & Planck2018 TT,TE,EE+lowE \quad\quad& Planck2018 TT,TE,EE+lowE\quad\quad&Planck2018 TT,TE,EE+lowE\quad\quad&Planck2018 TT,TE,EE+lowE\\
& +lensing+BK\textbf{14}+BAO &
+lensing+BK\textbf{14}+BAO&lensing+BK\textbf{18}+BAO&lensing+BK\textbf{18}+BAO\\
\hline \\$\lambda$& $68\%$ CL & $95\%$ CL &$68\%$ CL & $95\%$ CL\\
\hline\hline \\  $0.4$ & $-1.090<n<-0.998$ & $-1.134<n<-0.970$
& $\textsf{Not\, Consistent}$  & $-1.114<n<-1.004$\\ \\
\hline
\\ $0.8$ & $-0.912<n<-0.835$ & $-0.947<n<-0.810$
& $\textsf{Not\, Consistent}$  & $-0.931<n<-0.840$
\\ \\ \hline\\
$1.2$ & $-0.823<n<-0.755$ & $-0.857<n<-0.733$
& $\textsf{Not\, Consistent}$  & $-0.842<n<-0.759$ \\ \\
\hline \hline
\end{tabular}
\end{center}
\end{table*}

\begin{table*} \tiny\tiny\caption{\small{\label{tab2} Ranges
of the parameter $n$ for some sample values of $\lambda$ in which
the tensor-to-scalar ratio and scalar spectral index of our model
with $Q=\alpha\phi^{n}$ and exponential potential are consistent
with different data sets with $\textbf{N=60}$.}}
\begin{center}
\tabcolsep=0.05cm\begin{tabular}{cccccc}
\\ \hline \hline \\ & Planck2018 TT,TE,EE+lowE \quad\quad& Planck2018 TT,TE,EE+lowE\quad\quad&Planck2018 TT,TE,EE+lowE\quad\quad&Planck2018 TT,TE,EE+lowE\\
& +lensing+BK\textbf{14}+BAO &
+lensing+BK\textbf{14}+BAO&lensing+BK\textbf{18}+BAO&lensing+BK\textbf{18}+BAO\\
\hline \\$\lambda$& $68\%$ CL & $95\%$ CL &$68\%$ CL & $95\%$ CL\\
\hline\hline \\  $0.8$ & $-1.065<n<-0.975$ & $-1.107<n<-0.946$
& $\textsf{Not\, Consistent}$  & $-1.088<n<-0.980$\\ \\
\hline
\\ $0.9$ & $-0.892<n<-0.815$ & $-0.927<n<-0.792$
& $\textsf{Not\, Consistent}$  & $-0.911<n<-0.820$
\\ \\ \hline\\
$1.0$ & $-0.808<n<-0.740$ & $-0.841<n<-0.718$
& $\textsf{Not\, Consistent}$  & $-0.826<n<-0.744$ \\ \\
\hline \hline
\end{tabular}
\end{center}
\end{table*}

\subsubsection{Power-Law Potential}

With  $Q=\alpha\phi^{n}$ and a power-law potential as Eq.~\eqref{eq67}, the density and tensor
perturbations are obtained as follows
\begin{eqnarray}
A_s^2=\frac {{\kappa}^{6}{b}^{2} \left( n+1 \right) ^{2} \left(
2\,n+1 \right) ^{2} \left( 1+\alpha\,{\phi}^{n} \right)
^{2}{\phi}^{2}}{48{\pi}^{2}\Big[
-\frac{3}{2}\,n{\alpha}^{2}{\kappa}^{2} \left( n+1 \right)
{\phi}^{2+2\,n}+ \left(
-3\,{\phi}^{n+2}\alpha\,{\kappa}^{2}n+{b}^{2} \left( n+1\right)
\right)  \left( n+\frac{1}{2} \right)\Big] ^{2}}\\\nonumber
\times{{\rm exp}\bigg\{-{\frac {9\alpha{\kappa}^{2} \big( \alpha\,
\left( n+2\right) {\phi}^{2+2\,n}+2\,{\phi}^{n+2} \left( n+1 \right)
\big) }{2b \left( n+2 \right)  \left( n+1 \right)}}\bigg\}}\,,\hspace{3cm}
\end{eqnarray}
	
\begin{eqnarray}
A_T^2={\frac {4\,{\kappa}^{4}\beta\,{\phi}^{b}}{75\,\pi}{{\rm
exp}\bigg\{-\,{\frac {3\alpha{\kappa}^{2}
\big( \alpha\, \left( n+2 \right) {\phi}^{2+2\,n}+2\,{\phi}^{n+2} \left( n+1 \right)
\big) }{2b \left( n+2\right)  \left( n+1 \right) }}\bigg\}}}\,.\hspace{2.5cm}
\end{eqnarray}
	
To study the model numerically, we consider both quadratic ($b=2$)
and quartic ($b=4$) potentials. In this way, we obtain
observationally viable ranges of $\alpha$ and $n$ with both
potentials and in confrontation with both Planck2018 TT, TE, and
EE+lowE+lensing+BAO+BK\textbf{14} and Planck2018 TT, TE, and
EE+lowE+lensing+BAO+BK\textbf{18} datasets with e-folds numbers $N=55$ and $N=60$.
The results are shown in figure 9 and 10 respectively.

The $r-n_{s}$ plane with both potentials for $N=55$ and $N=60$ and in the background
of the mentioned datasets are shown in figures 11, 12, 13 and 14 respectively. The
constraints obtained in these cases are summarized in tables 5, 6, 7 and 8 respectively.

\begin{figure}
\begin{center}{\includegraphics[width=.4 \textwidth,origin=c,angle=0]{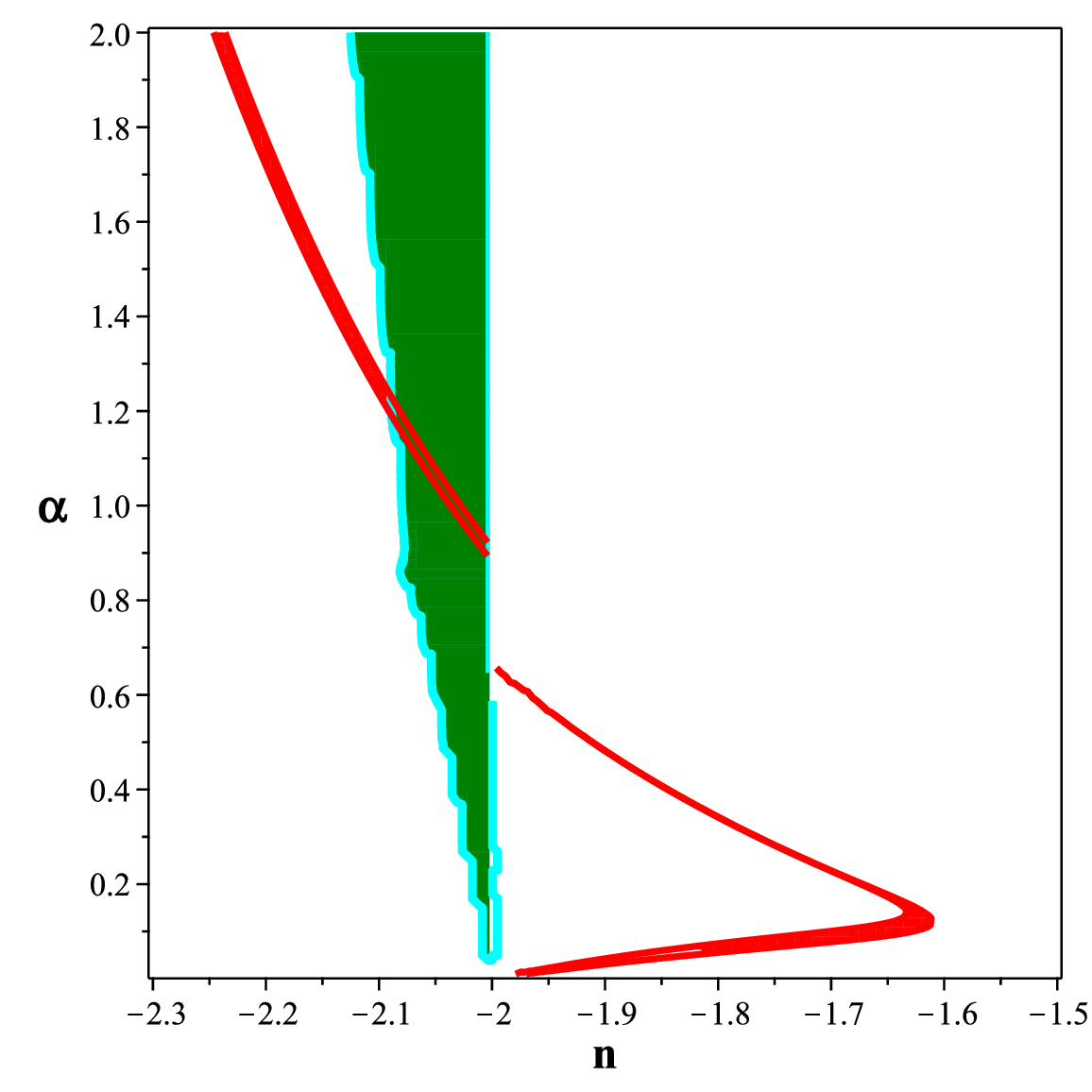}}
{\includegraphics[width=.4 \textwidth,origin=c,angle=0]{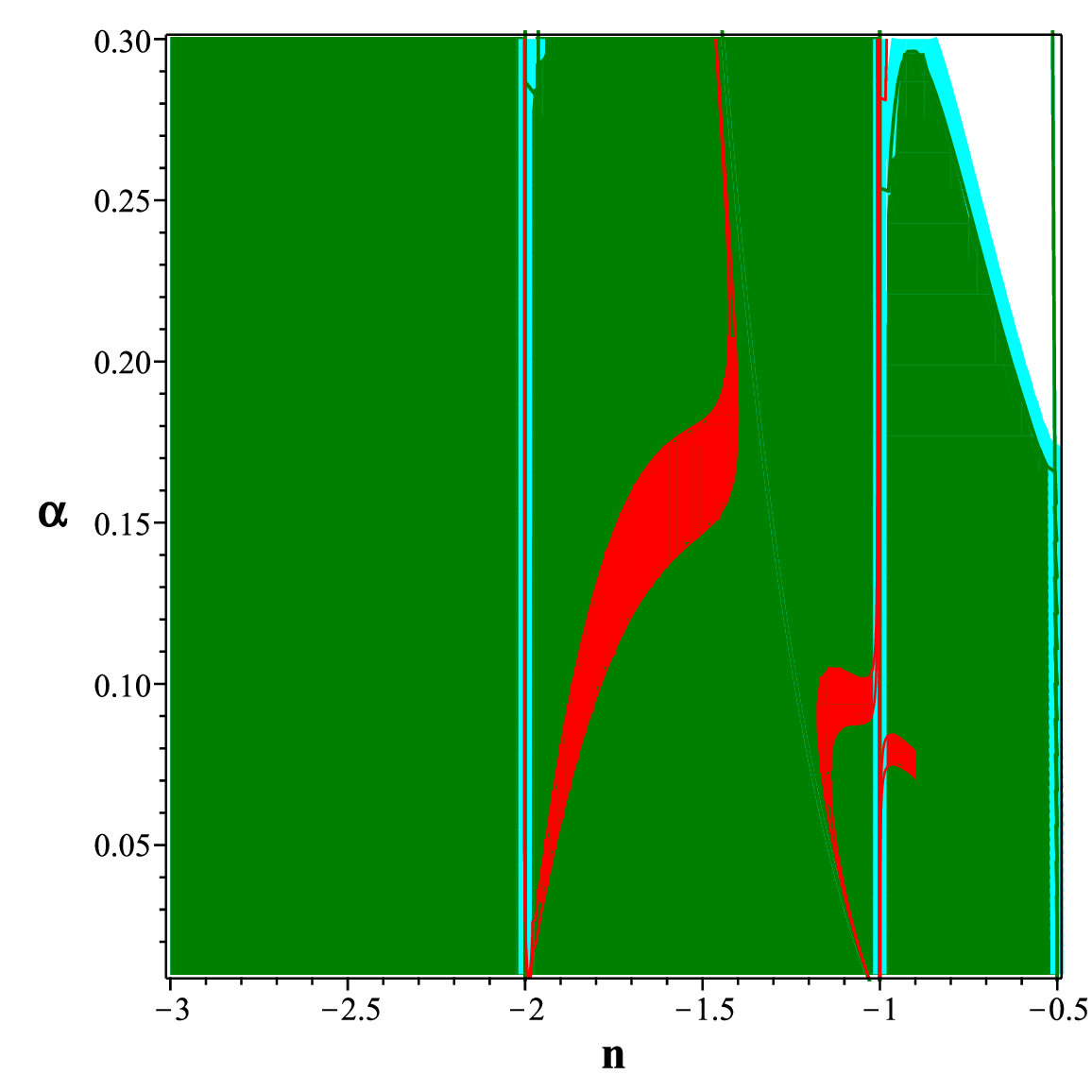}}	
\end{center}
\caption{\label{fig5}\small {Ranges of the parameters $\alpha$ and
$n$ with $\textbf{N=55}$ leading to observationally viable values of $r$ and $n_{s}$
with $Q=\alpha\phi^{n}$ and quadratic potential (left
panel), and also $Q=\alpha\phi^{n}$ and quartic potential (right
panel). The red region shows the values of the parameters leading to
observationally viable values of $n_{s}$ in confrontation with
Planck2018 TT, TE, and EE+lowE+lensing+BAO+BK\textbf{14} data. The cyan
region shows the values of the parameters leading to observationally
viable values of $r$ in confrontation with Planck2018 TT, TE, and
EE+lowE+lensing+BAO+BK\textbf{14} data. The green region shows the values of
the parameters leading to observationally viable values of $r$ in
confrontation with Planck2018 TT, TE, and EE+lowE+lensing+BAO+BK\textbf{18}
data.}}
\end{figure}
	
\begin{figure}
\begin{center}{\includegraphics[width=.4 \textwidth,origin=c,angle=0]{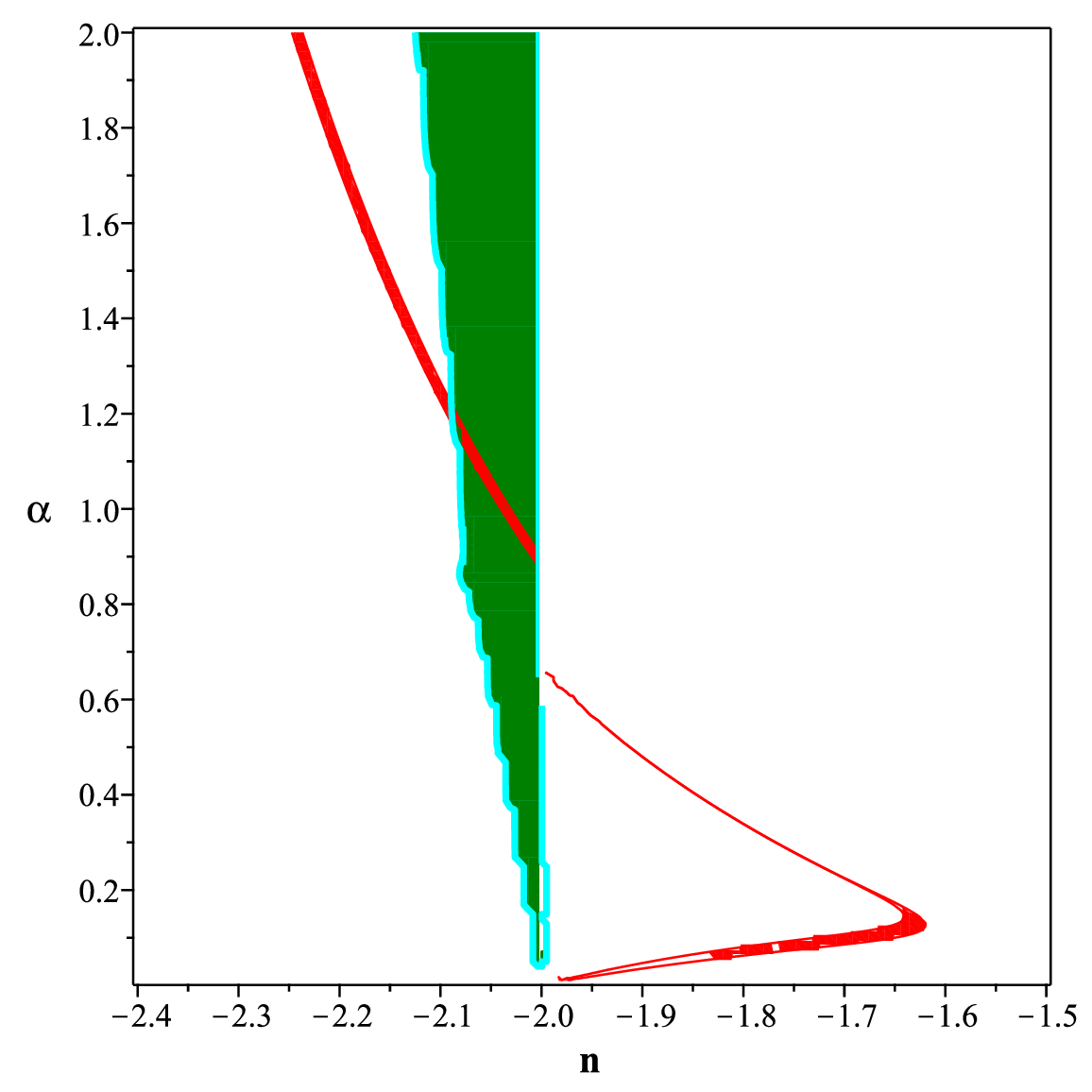}}
{\includegraphics[width=.4 \textwidth,origin=c,angle=0]{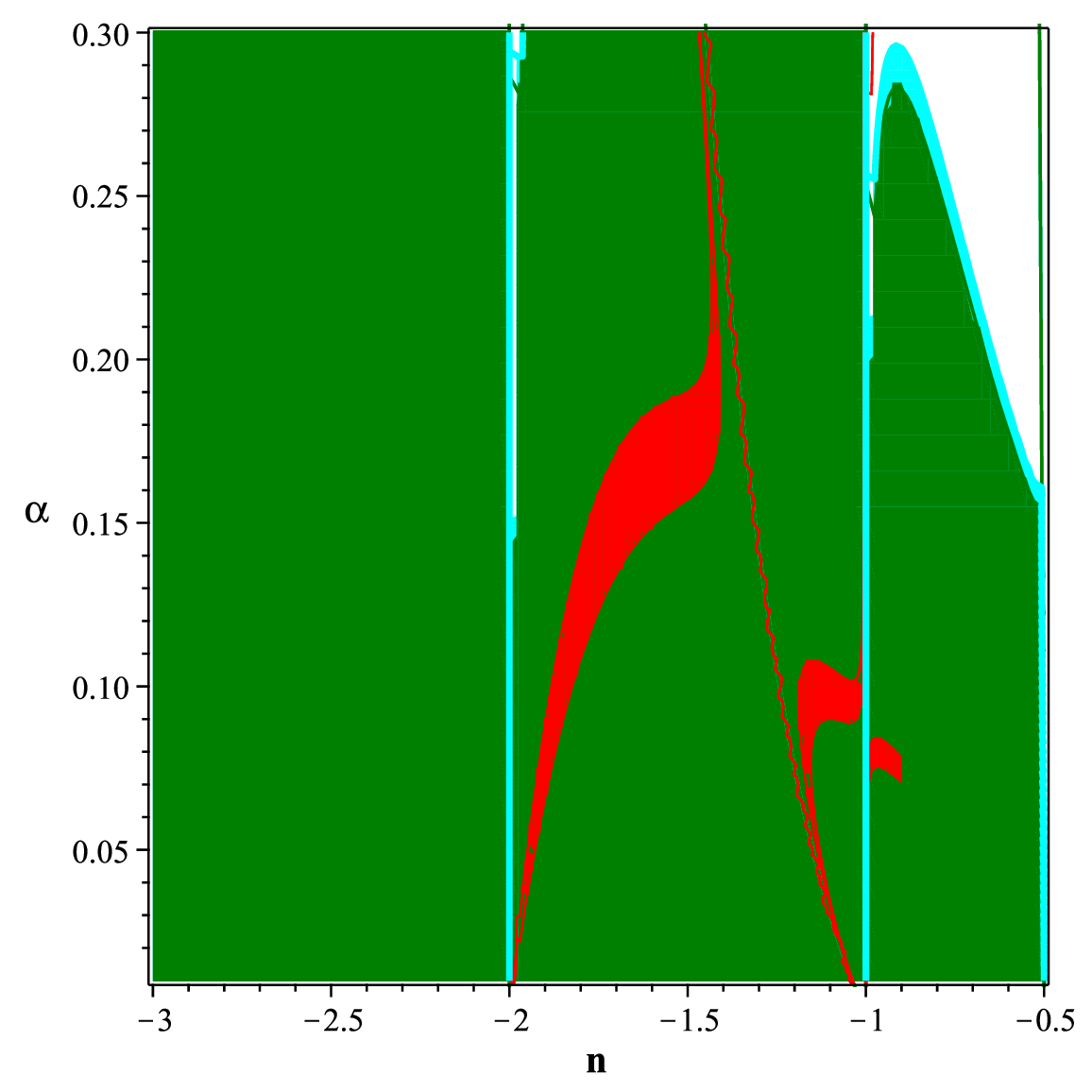}}	
\end{center}
\caption{\label{fig5}\small {Ranges of the parameters $\alpha$ and
$n$ with $\textbf{N=60}$ leading to observationally viable values of $r$ and $n_{s}$
with $Q=\alpha\phi^{n}$ and quadratic potential (left
panel), and also $Q=\alpha\phi^{n}$ and quartic potential (right
panel). The red region shows the values of the parameters leading to
observationally viable values of $n_{s}$ in confrontation with
Planck2018 TT, TE, and EE+lowE+lensing+BAO+BK\textbf{14} data. The cyan
region shows the values of the parameters leading to observationally
viable values of $r$ in confrontation with Planck2018 TT, TE, and
EE+lowE+lensing+BAO+BK\textbf{14} data. The green region shows the values of
the parameters leading to observationally viable values of $r$ in
confrontation with Planck2018 TT, TE, and EE+lowE+lensing+BAO+BK\textbf{18}
data.}}
\end{figure}

\begin{figure}
\begin{center}{\includegraphics[width=.66 \textwidth,origin=c,angle=0]{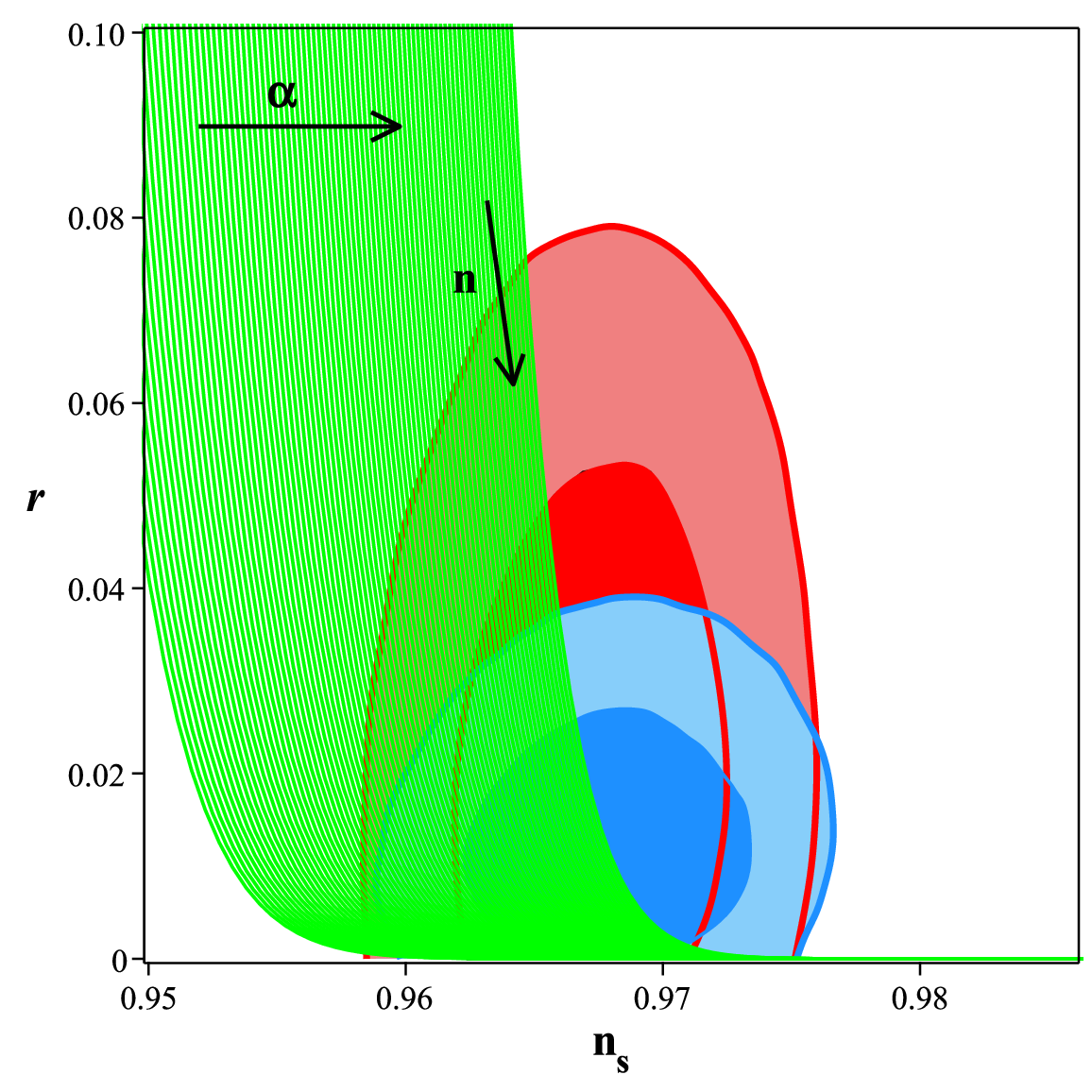}}
\end{center}
\caption{\label{fig6}\small {The $r-n_{s}$ plane with $Q=\alpha\phi^{n}$ and quadratic potential (green region) in the background
of Planck2018 TT, TE, and EE+lowE+lensing+BAO+BK\textbf{14} (red region) and
Planck2018 TT, TE, and EE+lowE+lensing+BAO+BK\textbf{18} (blue region) datasets with $\textbf{N=55}$. The ranges of $\alpha$ and $n$ are as $0.8< \alpha <1.4$ and $-3.0< n <-2.0$.}}
\end{figure}

\begin{figure}
\begin{center}{\includegraphics[width=.66 \textwidth,origin=c,angle=0]{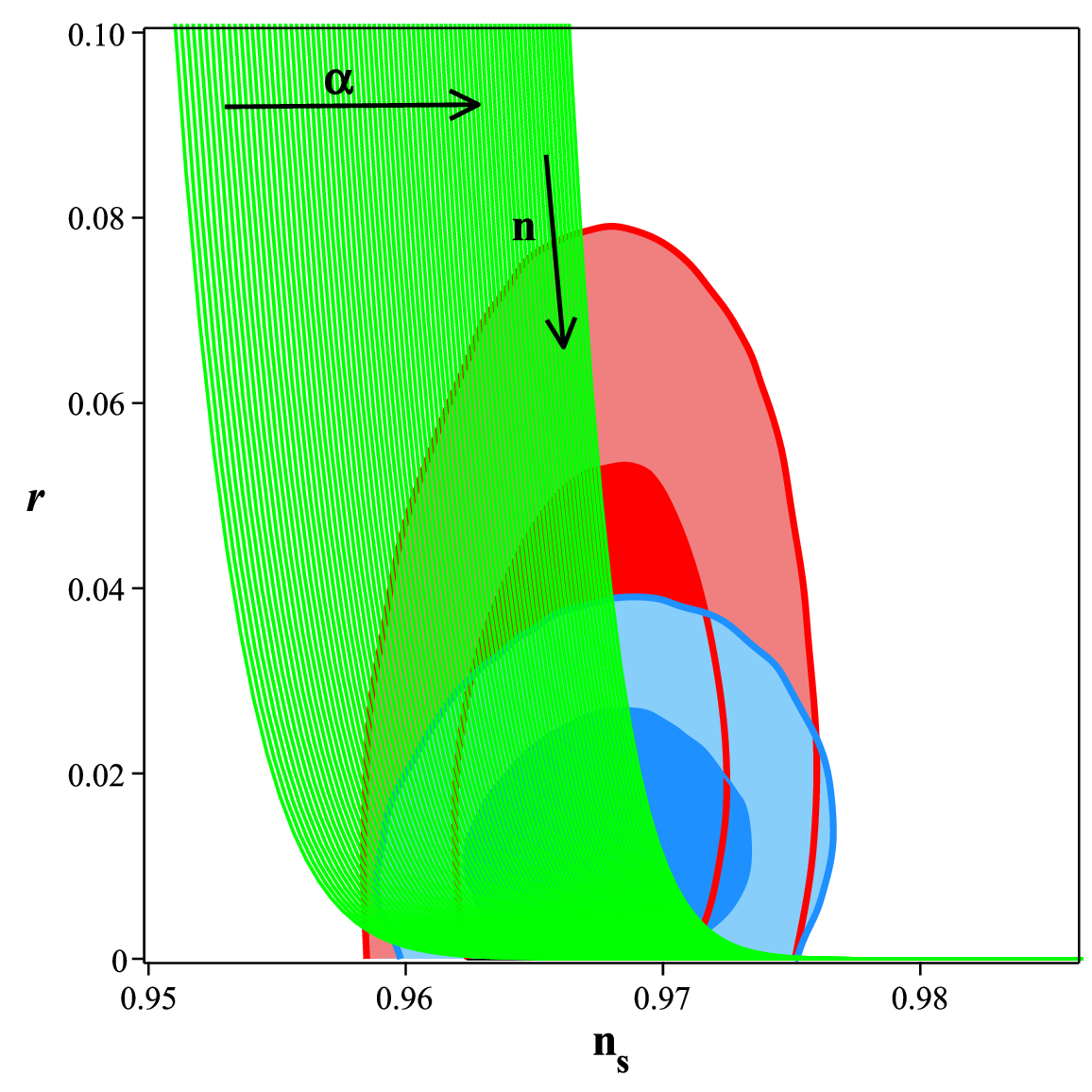}}
\end{center}
\caption{\label{fig6}\small {The $r-n_{s}$ plane with $Q=\alpha\phi^{n}$ and quadratic potential (green region) in the background
of Planck2018 TT, TE, and EE+lowE+lensing+BAO+BK\textbf{14} (red region) and
Planck2018 TT, TE, and EE+lowE+lensing+BAO+BK\textbf{18} (blue region) datasets with $\textbf{N=60}$. The ranges of $\alpha$ and $n$ are as $0.8< \alpha <1.4$ and $-3.0< n <-2.0$.}}
\end{figure}

\begin{figure}
\begin{center}{\includegraphics[width=.66 \textwidth,origin=c,angle=0]{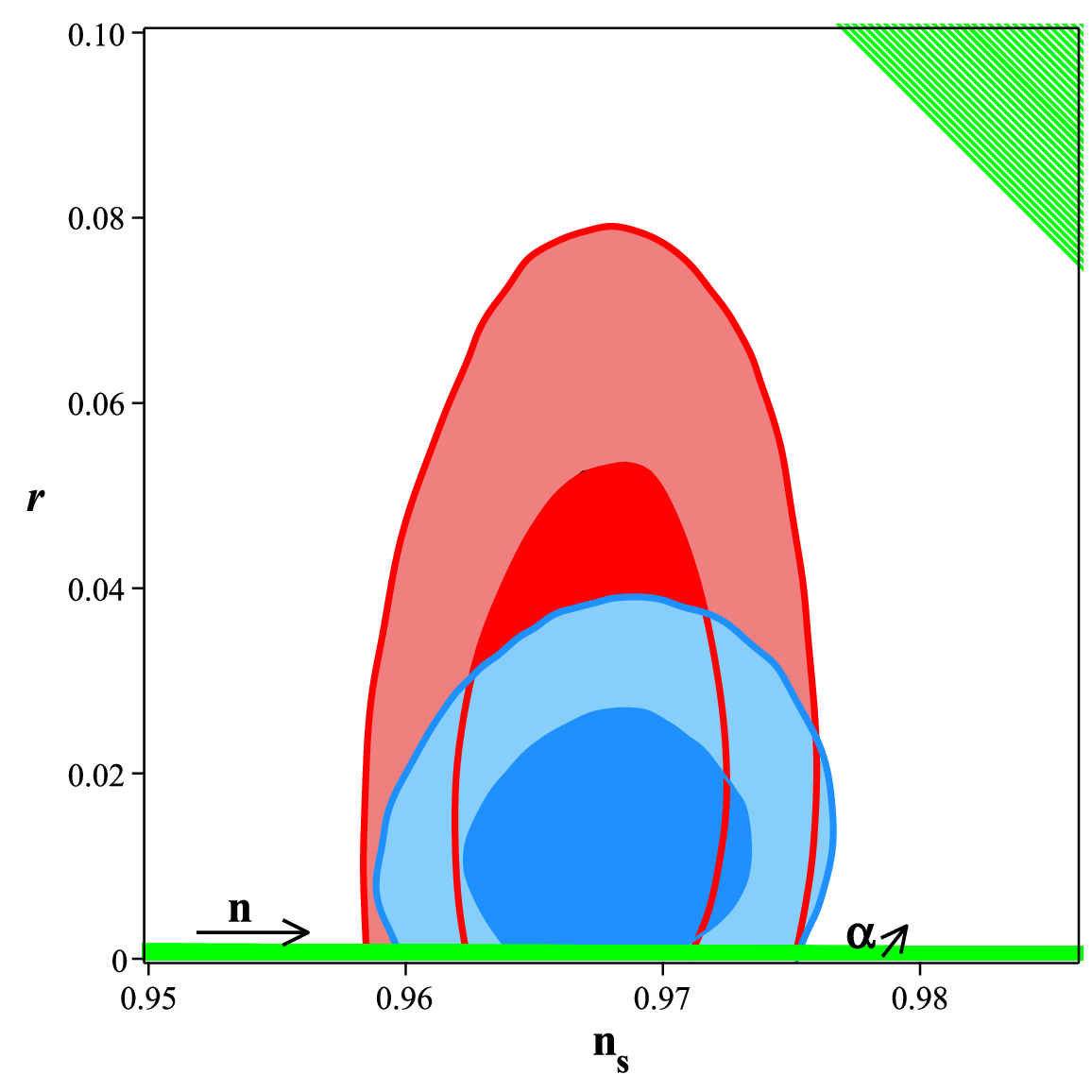}}
\end{center}
\caption{\label{fig7}\small {The $r-n_{s}$ plane with $Q=\alpha\phi^{n}$ and quartic potential (green region) in the background of
Planck2018 TT, TE, and EE+lowE+lensing+BAO+BK\textbf{14} (red region) and Planck2018
TT, TE, and EE+lowE+lensing+BAO+BK\textbf{18} (blue region) datasets with $\textbf{N=55}$. The ranges of $\alpha$ and $n$ are as $0.6< \alpha <1.2$ and $-2.0< n <-1.0$.}}
\end{figure}

\begin{figure}
\begin{center}{\includegraphics[width=.66 \textwidth,origin=c,angle=0]{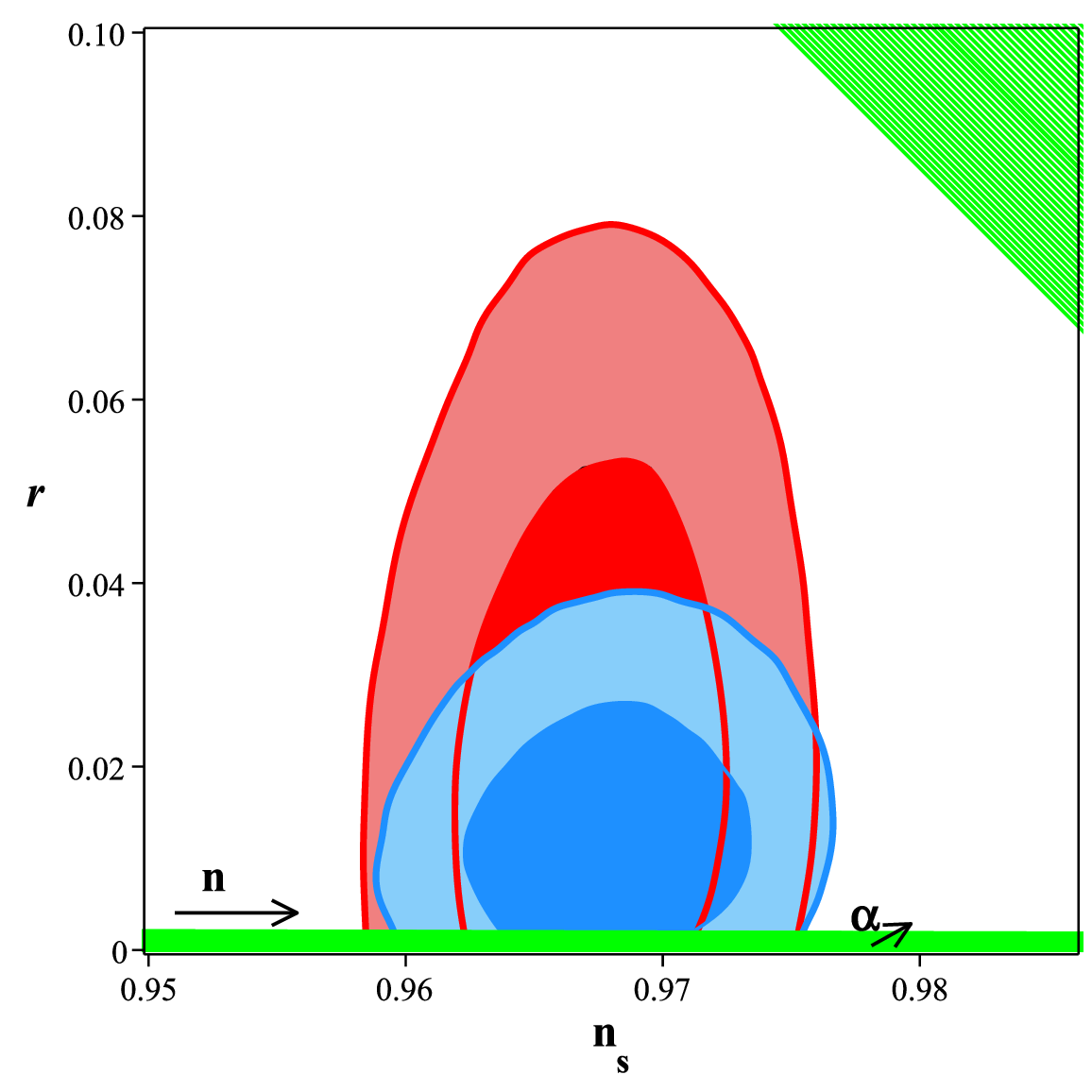}}
\end{center}
\caption{\label{fig7}\small {The $r-n_{s}$ plane with $Q=\alpha\phi^{n}$ and quartic potential (green region) in the background of
Planck2018 TT, TE, and EE+lowE+lensing+BAO+BK\textbf{14} (red region) and Planck2018
TT, TE, and EE+lowE+lensing+BAO+BK\textbf{18} (blue region) datasets with $\textbf{N=60}$. The ranges of $\alpha$ and $n$ are as $0.6< \alpha <1.2$ and $-2.0< n <-1.0$.}}
\end{figure}

\begin{table*} \tiny\tiny\caption{\small{\label{tab3} Ranges
of the parameter $n$ for some sample values of $\alpha$ in which
the tensor-to-scalar ratio and scalar spectral index with $Q=\alpha\phi^{n}$ and quadratic potential are consistent with
different datasets with $\textbf{N=55}$.}}
\begin{center}
\tabcolsep=0.06cm\begin{tabular}{cccccc}
\\ \hline \hline \\ & Planck2018 TT,TE,EE+lowE\quad\quad & Planck2018 TT,
TE,EE+lowE\quad\quad&Planck2018 TT,TE,EE+lowE\quad\quad&Planck2018 TT,TE,EE+lowE\\
& +lensing+BK\textbf{14}+BAO &
+lensing+BK\textbf{14}+BAO&lensing+BK\textbf{18}+BAO&lensing+BK\textbf{18}+BAO\\
\hline \\$\alpha$& $68\%$ CL & $95\%$ CL &$68\%$ CL & $95\%$ CL\\
\hline\hline \\  $1.2$ & $-2.087<n<-2.079$ & $-2.089<n<-2.074$
& $-2.0857<n<-2.079$  & $-2.0865<n<-2.073$\\ \\
\hline
\\ $1.1$ & $-2.065<n<-2.054$ & $-2.069<n<-2.048$
& $-2.063<n<-2.054$  & $-2.068<n<-2.048$
\\ \\ \hline\\
$1.0$ & $-2.037<n<-2.026$ & $-2.041<n<-2.020$
& $-2.034<n<-2.027$  & $-2.0394<n<-2.020$ \\ \\
\hline \hline
\end{tabular}
\end{center}
\end{table*}

\begin{table*} \tiny\tiny\caption{\small{\label{tab3} Ranges
of the parameter $n$ for some sample values of $\alpha$ in which
the tensor-to-scalar ratio and scalar spectral index with $Q=\alpha\phi^{n}$ and quadratic potential are consistent with
different datasets with $\textbf{N=60}$.}}
\begin{center}
\tabcolsep=0.06cm\begin{tabular}{cccccc}
\\ \hline \hline \\ & Planck2018 TT,TE,EE+lowE\quad\quad & Planck2018 TT,
TE,EE+lowE\quad\quad&Planck2018 TT,TE,EE+lowE\quad\quad&Planck2018 TT,TE,EE+lowE\\
& +lensing+BK\textbf{14}+BAO &
+lensing+BK\textbf{14}+BAO&lensing+BK\textbf{18}+BAO&lensing+BK\textbf{18}+BAO\\
\hline \\$\alpha$& $68\%$ CL & $95\%$ CL &$68\%$ CL & $95\%$ CL\\
\hline\hline \\  $1.0$ & $-2.039<n<-2.029$ & $-2.044<n<-2.023$
& $-2.038<n<-2.030$  & $-2.042<n<-2.023$\\ \\
\hline
\\ $1.1$ & $-2.068<n<-2.057$ & $-2.072<n<-2.051$
& $-2.066<n<-2.058$  & $-2.070<n<-2.050$
\\ \\ \hline\\
$1.2$ & $-2.087<n<-2.081$ & $-2.088<n<-2.076$
& $-2.085<n<-2.081$  & $-2.086<n<-2.076$ \\ \\
\hline \hline
\end{tabular}
\end{center}
\end{table*}

\begin{table*} \tiny\tiny\caption{\small{\label{tab1} Ranges
of the parameter $n$ for some sample values of $\alpha$ in which
the tensor-to-scalar ratio and scalar spectral index with $Q=\alpha\phi^{n}$ and quartic potential are consistent with
different datasets with $\textbf{N=55}$.}}
\begin{center}
\tabcolsep=0.05cm\begin{tabular}{cccccc}
\\ \hline \hline \\ & Planck2018 TT,TE,EE+lowE \quad\quad& Planck2018 TT,TE,EE+lowE\quad\quad&Planck2018 TT,TE,EE+lowE\quad\quad&Planck2018 TT,TE,EE+lowE\\
& +lensing+BK\textbf{14}+BAO &
+lensing+BK\textbf{14}+BAO&lensing+BK\textbf{18}+BAO&lensing+BK\textbf{18}+BAO\\
\hline \\$\alpha$& $68\%$ CL & $95\%$ CL &$68\%$ CL & $95\%$ CL
\\
\hline\hline \\  $0.8$ & $-1.01949<n<-1.01944$ &
$-1.01951<n<-1.01942$
& $-1.01948<n<-1.01945$  & $-1.01950<n<-1.01943$\\ \\
\hline
\\  $0.9$ & $-1.02150<n<-1.02146$ & $-1.02151<n<-1.02144$
& $-1.02149<n<-1.02146$  & $-1.02150<n<-1.02144$
\\ \\ \hline\\
$1.0$ & $-1.02348<n<-1.02345$ & $-1.02349<n<-1.02343$
& $-1.02347<n<-1.02345$  & $-1.02349<n<-1.02343$ \\ \\
\hline \hline
\end{tabular}
\end{center}
\end{table*}

\begin{table*} \tiny\tiny\caption{\small{\label{tab1} Ranges
of the parameter $n$ for some sample values of $\alpha$ in which
the tensor-to-scalar ratio and scalar spectral index with $Q=\alpha\phi^{n}$ and quartic potential are consistent with
different datasets with $\textbf{N=60}$.}}
\begin{center}
\tabcolsep=0.05cm\begin{tabular}{cccccc}
\\ \hline \hline \\ & Planck2018 TT,TE,EE+lowE \quad\quad& Planck2018 TT,TE,EE+lowE\quad\quad&Planck2018 TT,TE,EE+lowE\quad\quad&Planck2018 TT,TE,EE+lowE\\
& +lensing+BK\textbf{14}+BAO &
+lensing+BK\textbf{14}+BAO&lensing+BK\textbf{18}+BAO&lensing+BK\textbf{18}+BAO\\
\hline \\$\alpha$& $68\%$ CL & $95\%$ CL &$68\%$ CL & $95\%$ CL
\\
\hline\hline \\  $0.8$ & $-1.01863<n<-1.01859$ &
$-1.01865<n<-1.01857$
& $-1.01862<n<-1.01859$  & $-1.01864<n<-1.01857$\\ \\
\hline
\\  $0.9$ & $-1.02056<n<-1.02053$ & $-1.02058<n<-1.02053$
& $-1.02055<n<-1.02053$  & $-1.02057<n<-1.02051$
\\ \\ \hline\\
$1.0$ & $-1.02247<n<-1.02244$ & $-1.02249<n<-1.02243$
& $-1.02247<n<-1.02244$  & $-1.02248<n<-1.02242$ \\ \\
\hline \hline
\end{tabular}
\end{center}
\end{table*}

Before coming to the end, we stress on an important issue regarding the dissipative
quintessence as a potential dark energy candidate. It is well known that a non-dissipative quintessence field
can be a dark energy candidate driving the late time cosmic expansion if its
equation of state parameter to be in the range $-1/3<w<-1$, where $w=P/\rho$. About a dissipative
quintessence as a potential candidate for the dark energy, it is important to note
that dissipation for a quintessence field is essentially considerable in early universe where
the density, pressure and temperature for the cosmic fluid were moderate. This is the reason why
dissipation in our setup is devoted to early time inflationary expansion of the universe. For
the late time cosmic evolution, dissipation of the quintessence field can be neglected easily
due to low density, low pressure and low temperature of the cosmic fluid. Therefore, for a small
dissipation, the well-known existing arguments for a non-dissipative quintessence as the dark energy candidate works well.

\section{Summary and Conclusion}

In this paper we have constructed a cosmological inflation model within a dissipative
quintessence framework. We have presented a Lagrangian formulation of the dissipative
system whose theoretical description can be obtained from a variational
principle. Then we have studied the inflationary dynamics and dissipation
effects on the model parameters space. The inflation parameters and perturbations have been
calculated in detail. We have considered power-law and
exponential potentials as some ansatz to obtain scalar spectral index and
tensor-to-scalar ratio for a constant as well as variable dissipation
factor. Depending on the scalar field potential model and the form of the dissipation function,
there are different behaviors for the inflation parameters. In this regard we have explored
the evolution of the inflationary parameters in confrontation with the recent joint
observational data with two different numbers of e-folds; $N=55$ and $N=60$. Then we have obtained
some constraints on the model's parameters space which have led to the values of the scalar spectral index and tensor-to-scalar
ratio consistent with $68\%$ and $95\%$ confidence levels of the Planck2018
TT, TE, EE+lowE+lensing+BAO+BK\textbf{14} and Planck2018 TT, TE, EE+lowE+lensing+BAO +BK\textbf{18} data.
According to our analysis, for exponential potential with both cases of constant and variable dissipation
functions, the $r-n_s$ results with $N=55$ are consistent with  Planck2018 TT, TE, EE+lowE+lensing
data at the $68\%$ CL and $95\%$ CL for some ranges of the parameters $\lambda$, $Q_0$ and $n$.
However, this consistency for the constant dissipation and $N=60$ is mildly with small and negative dissipation
factor and negligible tensor-to-scalar ratio. In this respect, the dissipative quintessential inflation model with $N=55$
is observationally more viable than the case with $N=60$. For the variable dissipation factor,
the consistency with observation for both $N=55$ and $N=60$ is more or less the same (figures 7 and 8),
with more acceptable tensor-to-scalar ratio. The quadratic and quartic potentials with variable
dissipation function are consistent with Planck2018 TT, TE, EE+lowE+lensing
data at the $68\%$ CL and $95\%$ levels of confidence for some ranges of the parameters $\alpha$ and $n$.
A comparison between the cases with and without dissipation (that is, $Q\neq 0$ and $Q=0$)
in our setup shows that a quintessential inflation with, for instance, power-law potential is more consistent with observation.
In fact, our treatment in this paper reveals that a dissipative quintessential inflation is more reliable from observational viewpoint than the
standard non-dissipative quintessential inflation with the same adopted potentials and in the same level of confidence.\\

{\bf Acknowledgement}
We appreciate the contribution of an anonymous referee for very constructive and insightful comments.

\end{document}